%% file: main.tex
\documentclass[10.75pt, a4paper]{article}
\input{initialisation.tex}
\usepackage[framemethod=tikz]{mdframed}
\usepackage{lipsum}
\usepackage{dirtytalk}
\usepackage[most]{tcolorbox}
\usepackage[font=footnotesize]{caption}
\newtcolorbox{mybox}[1]{colback=green!6!white,colframe=black!75!black,fonttitle=\bfseries,title=#1}
\newtcolorbox{mybox2}{colback=red!5!white,colframe=red!75!black}

\usepackage{pifont}

\newtcolorbox{myBox}[3][]{
arc=2mm,
lower separated=true,
fonttitle=\bfseries,
colbacktitle=gray!10,
coltitle=black!50!black,
colframe=gray!10,
colback=gray!10,
title=#2,#1}

\usepackage{dirtytalk}
\usepackage[font=footnotesize]{caption}

\usepackage{pifont}

\usepackage{soul,xcolor}
\setstcolor{red}

\usepackage{xcolor,hyperref}
\hypersetup{
   colorlinks,
   linkcolor={blue!50!black},
   citecolor={blue!50!black},
   urlcolor={blue!80!black}
} 

\definecolor{mycolor}{rgb}{0.122, 0.435, 0.698}

\newcommand{\Ubold}{\textbf{u}}
\newcommand{\del}[2]{\frac{\partial#1}{\partial#2}}
\newcommand{\even}{{\mathcal{E}}}
\newcommand{\odd}{{\mathcal{O}}}

\title{Odd Droplets: \protect\\
Fluids with Odd Viscosity and Highly Deformable Interfaces}
\author[1,2]{Hugo França\footnote{h.franca@uva.nl, ORCID: 0000-0002-5361-7704}}
\author[2]{Maziyar Jalaal\footnote{m.jalaal@uva.nl, ORCID: 0000-0002-5654-8505}}

\affil[1]{Advanced Research Center for Nanolithography, Science Park 106, 1098 XG Amsterdam, The Netherlands}
\affil[2]{Van der Waals-Zeeman Institute, Institute of Physics, University of Amsterdam,\protect\\
Science Park 904, Amsterdam, 1098XH, The Netherlands}

\begin{document}
\input{colors_define.tex}
\begingroup
\sffamily
\date{}
\maketitle
\endgroup


\begin{abstract}

Flows with deformable interfaces are commonly controlled by applying an external field or modifying the boundaries that interact with the fluid, but realizing such solutions can be demanding or impractical in various scenarios. Here, we demonstrate that fluids with broken symmetries can self-control their mechanics.
We present a continuum model of a viscous fluid with highly deformable interfaces subject to capillary stresses. Our model features odd viscosity, a parity-violating property that emerges in chiral fluids. Using direct numerical simulations, we focus on the impact of an odd droplet on a superhydrophobic surface. We demonstrate that odd viscosity dramatically disrupts conventional symmetric spreading by inducing asymmetric deformations and chiral flow patterns. Our analysis reveals a variety of dynamic regimes, including leftward and rightward bouncing, as well as rolling, depending on the relative strength of the odd viscosity. Our work illustrates that regulating odd viscosity provides a promising framework for controlling multiphase flows and designing functional metamaterials with tailored fluidic properties.


\textbf{keywords: Odd Viscosity $|$ Droplet Dynamics $|$ Active Matter $|$ Fluid Dynamics $|$ Chirality } 

\end{abstract}


\section*{Introduction} 

The principles of symmetry and symmetry breaking lie at the heart of understanding materials and phenomena across scales, both at equilibrium and far from equilibrium. Conventional fluids often exhibit time-reversal symmetry (TRS) at equilibrium. However, fluids like two-dimensional electron fluids in magnetic fields~\cite{avron1995viscosity,delacretaz2017transport,scaffidi2017hydrodynamic,berdyugin2019measuring,cosme2023nonlinear,fritz2024hydrodynamic} or chiral active fluids~\cite{furthauer2012active,banerjee2017odd,soni2019odd,markovich2021odd,bililign2022motile}, can break TRS, leading to non-dissipative, parity-violating odd viscosity~\cite{avron1998odd,Fruchart2023}. This unique property has been associated with various nonintuitive phenomena on different scales, such as Hall viscosity~\cite{delacretaz2017transport,berdyugin2019measuring,}, pattern formation in turbulent flows~\cite{de2024pattern,chen2024odd}, topological edge modes~\cite{souslov2019topological,tauber2020anomalous}, and anisotropic mobility, drag, and lift~\cite{hosaka2021hydrodynamic,khain2022stokes,lier2023lift,poggioli2023odd,aggarwal2023thermocapillary,lier2024odd}. Many such flows could naturally involve deformable interfaces~\cite{soni2019odd,
abanov2018odd,kirkinis2019odd,abanov2019free,samanta2022role,hosaka2021hydrodynamic, aggarwal2023thermocapillary}, where surface tension and stress boundary conditions give rise to rich, highly nonlinear behaviors. These shape-changing interfacial dynamics are intricately coupled with bulk flow and asymmetric stresses and are hardly accessible with analytical methods, particularly when deformations are large. Here, we present direct numerical simulations (DNS) of 2D flows with odd viscosity using a continuum model to study sharp deformable interfaces with surface tension, revealing unprecedented nonlinear responses and showing how odd viscosity can be harnessed as a parameter to control multiphase flows and design functional metamaterials~\cite{urzhumov2011fluid,park2019hydrodynamic,ewoldt2022designing,djellouli2024shell}. 

To this end, we study the rich problem of droplet dynamics and, in particular, droplet impact~\cite{josserand2016drop}. The impact of droplets on a substrate or fluid surface has long been a cornerstone of fluid dynamics, with classical studies revealing a wide range of phenomena depending on factors such as impact velocity, surface tension, viscosity, and wettability~\cite{yarin2006drop,josserand2016drop}. Traditional results include the spreading and retracting of droplets, crown-like splashing at high impact speeds, and intricate patterns formed by capillary waves. These dynamics are controlled by the balance of inertial, viscous, and surface tension forces, often characterized by dimensionless numbers such as Reynolds, Weber, and Ohnesorge numbers. 
In this letter, we focus on the impact of a 2D droplet on a non-wetting surface~\cite{clanet2004maximal,quere2005non,wildeman2016spreading,zhang2022impact} and show that introducing odd viscosity fundamentally alters these familiar scenarios: the parity-violating stress generates nontrivial flows, which deform the droplet interface. These deformations, in turn, modify the flow field, creating a self-reinforcing interplay between the interface shape and the underlying hydrodynamics. As we will discuss, this emergent two-way coupling leads to novel nonlinear behaviors, such as asymmetric spreading, asymmetric bouncing, and rolling, all of which are inaccessible in conventional fluids.


\section*{Problem Statement}


Consider a 2D droplet of an odd viscous fluid $\Omega_1$ with diameter $D_0$ and constant density $\rho_d$ impacting a surface with constant velocity $U$ at $t=0$. The surrounding fluid $\Omega_2$ is Newtonian and has a much smaller density and (even) viscosity than the droplet. We solve an extension of the Navier–Stokes equations for an incompressible fluid (see Appendix~\ref{appendix:Equations} for details of the equations, boundary conditions, and nondimensionalization):


\begin{gather} 
\nabla \cdot \bm{u} = 0 \\
\mathrm{D}\,\bm{u}\, /\, \mathrm{D}\, t 	= - \bm{\nabla} p  + 
\bm{\nabla} \cdot \bm{\tau}^\even + \bm{F}^{\mathcal{O}}  + \bm{F}^{\mathcal{S}}
\label{eq:nondim_navier}
\end{gather}


Here, $\mathrm{D}\,\bm{u}\, /\, \mathrm{D}\, t = \partial /\partial \, t + \bm{u} \cdot \bm{\nabla}$ is the material derivative, and $\bm{u} (\bm{x},t)$ and $p(\bm{x},t)$ are the velocity and pressure fields, respectively. 
The second term on the right-hand side is the divergence of the  deviatoric stress tensor $\bm{\nabla} \cdot \bm{\tau}^\even = \left( 
\bm{\nabla} \cdot  \dot{\bm{\bm{\gamma}}}^\even \right) / Re^\even$ and accounts for the dissipative even viscosity $\mu^\even$ in even strain rate field $\dot{\bm{\gamma}}^\even$, with the even Reynolds number $Re^\even = \rho_d \, U \, D_0 / \mu^\even$ measuring the inertial effects. 
The odd effects arise from a force exhibiting the same structure with $\bm{F}^\odd \equiv \bm{\nabla} \cdot \bm{\tau}^\odd = \left( \bm{\nabla} \cdot  \dot{\bm{\gamma}}^\odd \right) / Re^\odd \sim - \bm{\nabla} \, \omega $, where $\omega(\bm{x},t)$ is the z-component of the vorticity field $ \bm{\omega}= \bm{\nabla} \times \bm{u}$, $\dot{\bm{\gamma}}^\odd$ is the odd strain rate field, and $Re^\odd = \rho_d \, U \, D_0 / \mu^\odd$ is the odd Reynolds number.  Nonetheless, the odd stress tensor, importantly, is not dissipative, \emph{i.e.}, $\bm{\tau}^\odd : \bm{\nabla} \bm{u} =0$~\cite{avron1995viscosity,Fruchart2023}. This means the odd stress tensor does not resist the flow and remains \say{reactive} to it. The last term in the right hand side of equation~\ref{eq:nondim_navier} accounts for surface tension forces, such that $\bm{F}^\mathcal{S} = \left(\kappa \, \delta_s \, \bm{n} \right) / We$. Here, $\kappa$ is the curvature of the interface between $\Omega_1$ and $\Omega_2$, and $\bm{n}$ is the normal vector to that interface. The Dirac function $\delta_s$ is centered on the interface, and the Weber number $We = \rho_d \, U^2 \,D_0 / \sigma$ compares the inertial and surface tension forces introduced by the surface tension $\sigma$. 
Gravity is ignored in all simulations presented here.
We use a combination of Finite Volume and Volume of Fluid (VoF) Methods~\cite{Hirt1981,popinet2009accurate,Popinet2015}, implemented in the open-source code Basilisk~\cite{Popinet2013-Basilisk} to solve these equations. To this end, a scalar color function $c(\bm{x},t)$ is introduced and convected by the droplet phase, $\Omega_1$. The densities and viscosities of the computational domain are hence, calculated as $\rho(c) =  c \ \rho_d + (1 - c)\rho_a$, $\mu^\even(c) = c \  \mu_d^\even + (1 - c)\mu_a^\even$, and\ $\mu^\odd(c) = c \  \mu_d^\odd$. Throughout this paper, we take the properties of the outer fluid as $\rho_a = 10^{-2}\rho_d$ and $\mu_a^\even = 10^{-2}\mu_d^\even$ (see Appendices~\ref{appendix:Equations} and \ref{appendix:numerical_implementation} for the details of equations, numerical implementation, and validations).

\section*{Odd Viscosity leads to chiral droplets with anomalous hydrodynamics}

\begin{figure}[t!]
\centering
\includegraphics[width=1\textwidth]{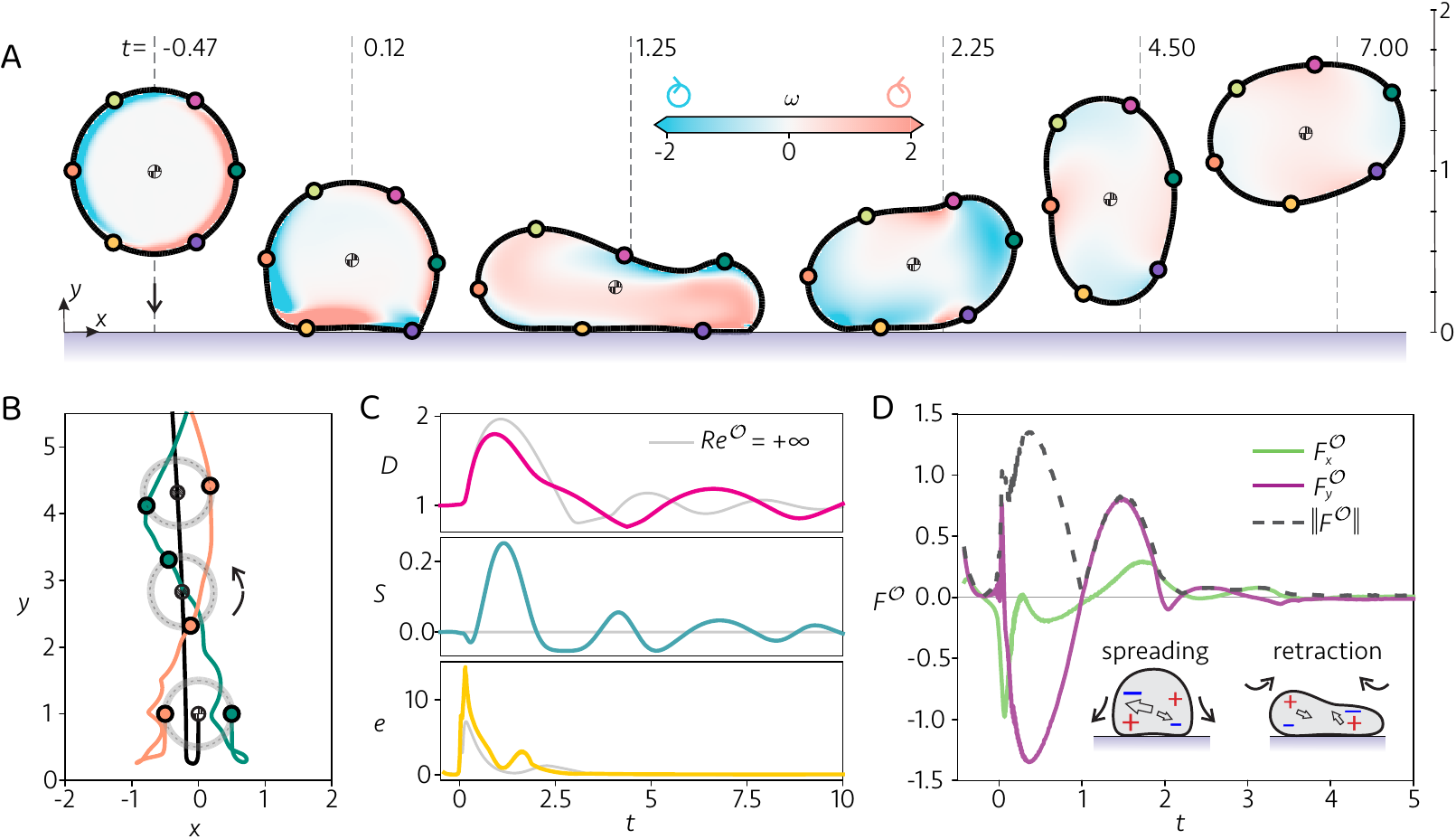}
\caption{\textbf{Anomalous dynamics of an odd droplet.} 
\textbf{A.} An impacting two-dimensional droplet with odd viscosity spreads asymmetrically on a rigid superhydrophobic surface and bounces back off-center while rotating. The color indicates the magnitude of the vorticity $\omega$. Lagrangian tracer particles placed on the droplet's surface highlight the surface flows and chirality.  
\textbf{B.} The center-of-mass trajectory of the droplet (black line) shows left/right symmetry breaking during spreading and bouncing. The droplet's rotation is visible by tracking two initially parallel tracer particles on its sides (orange and green lines).  
\textbf{C.} Characteristics of odd droplet impact manifest as a smaller spreading diameter $D$, significant shape asymmetry $S$, and bursts of enstrophy $e$, particularly upon impact. The gray lines show the results for the even counterpart, where no symmetry breaking is observed. \\ 
\textbf{D.} The odd force field $\bm{F}^{\mathcal{O}}$ evolves during both spreading and retraction in a non-trivial manner. In simplified terms, the vorticity quadrupole generated during spreading produces a net negative horizontal force. As retraction develops, the flow direction reverses, altering the odd effects (see Appendices~\ref{appendix:odd_force_distributions} for details of the flow and odd force fields). Note that the force direction is dictated by the sign of the odd viscosity, which is determined by microscopic chirality. At long times, the odd force decays to zero as the flow field becomes nearly purely rotational. See \textbf{Supplementary Video I} for a comparison of even and odd impacts.
}
\label{fig1:phenomenology}
\end{figure}

A Newtonian droplet impacting a rigid, non-wetting (superhydrophobic) surface spreads horizontally, dissipating energy through shear flow. Once it reaches maximum diameter ($D_{max}$), surface tension retracts the droplet, which may bounce fully or partially. Throughout, the droplet remains symmetric with respect to the axis of the fall. The introduction of odd viscosity drastically breaks this established picture. 
%
Figure \ref{fig1:phenomenology}A shows snapshots of the impact process with parameters $Re^\even = 50$, $Re^\odd = 10$ and $We = 10$ (also see Supplementary Video I). 
At these values, inertia, viscous dissipation, and odd viscous effects all play significant roles.
Prior to impact, the drag force from the surrounding fluid induces a small vorticity gradient within the droplet, giving rise to an odd force~\cite{hosaka2021hydrodynamic}. However, over the fall time considered in these simulations, these effects remain negligible.
The effects of odd viscosity appear immediately upon impact as the droplet spreads over the surface, with local vorticities near the solid boundary perturbing the flow field inside the droplet. 
This breaks the line symmetry in the flow field, resulting in an asymmetric shape and a leftward shift of the droplet’s center of mass (see Appendices~\ref{appendix:odd_force_distributions} and \ref{appendix:velocity_field_distributions} for a detailed analysis of the transient flow and odd force fields). 
During the transition from spreading to retraction, the flow anatomy becomes considerably more complex. The interplay of surface deformation and modulated bulk flows caused by odd viscosity produces a time-dependent, heterogeneous flow field within the droplet. Consequently, both the magnitude and direction of odd forces vary in space and time. Simply put, as retraction progresses, shear flow develops in the opposite direction to that during spreading, generating an odd force field that opposes the one generated during impact (see Appendix~\ref{appendix:odd_force_distributions} for a visualization of the odd force fields). 
This also initiates a chiral flow within the droplet and rotates it as it rebounds from the surface. In this case, the droplet's center-of-mass velocity drives a horizontal translation to the left as it ascends. Figure \ref{fig1:phenomenology}B shows an extended center-of-mass trajectory, highlighting the droplet’s counter-clockwise rotation and off-center bounce. 

We further quantify the effect of odd viscosity over time by three key properties (see Fig.~\ref{fig1:phenomenology}C): the droplet diameter ($D$), a shape symmetry parameter ($\mathcal{S}$), and the droplet enstrophy ($e = \int_{\Omega_1} \lVert \omega \rVert^2 $).
Odd droplets attain a smaller maximum diameter than their even counterpart, implying that the odd viscosity partly counteracts the droplet's spreading by generating flows that dissipate energy (also see Fig.~\ref{fig3:sweep_Re_We}A). At later times, further differences are visible as both capillary-driven oscillations and solid-body rotations induced by macroscopic chirality are present. 
The shape symmetry parameter, which quantifies the droplet's interface skewness~\cite{Joanes1998-Skewness}, is defined as $\mathcal{S} = \frac{1}{\Omega}\int_\Omega{(\bm{x} - \bm{x}_{CM})^3} / \left[ \frac{1}{\Omega} \int_\Omega{(\bm{x} - \bm{x}_{CM})^2} \right]^{3/2}$, where $\bm{x}_{CM}$ is the position vector of the droplet's center of mass. 
%
Odd viscosity introduces a strong loss of symmetry in the droplet shape, particularly immediately after impact when shear forces are most pronounced (see Fig.~\ref{fig1:phenomenology}C). 
This is also when the enstrophy is maximized. At later times, surface tension restores the shape symmetry by making the droplet circular again ($\mathcal{S} \rightarrow 0$, as $t \rightarrow \infty$), while both the flow’s chiral symmetry and the droplet’s translational symmetry remain broken, and enstrophy decays to nearly zero. 
To investigate the origins of the left/right symmetry breaking in more details, we examine the magnitude of the odd force acting inside the droplet, that is $\lVert \bm{F}^\odd \rVert= \frac{4}{\pi D_0^2}\int_\Omega\bm{F^\odd}$. This average force, along with its $x$ and $y$ components, is plotted over time in Fig.~\ref{fig1:phenomenology}D.
At the early stage of interaction with the wall, negative odd forces dominate, accelerating the droplet leftward on the surface; in contrast, as retraction develops, the odd forces can change sign, guiding the droplet rightward.
Note that the values shown in Fig.~\ref{fig1:phenomenology}D are spatially averaged over a heterogeneous odd force field arising from nonlinear interactions driven by inertia and modulated by odd viscosity, coupled with large surface deformations (see Appendix~\ref{appendix:odd_force_distributions}).
However, one may provide a simplified, instructive picture during spreading and retraction, where non-symmetric vorticity quadrupoles are formed (see Fig.~\ref{fig1:phenomenology}A and D and Appendix~\ref{appendix:odd_force_distributions}). 
Both left-right and top-bottom symmetries are broken in these quadrupoles, leading to net odd forces with horizontal components, leftward during spreading and rightward during retraction.
In the case depicted in Fig.~\ref{fig1:phenomenology}, the chiral droplet ultimately bounces leftward. However, discussed in the following, the phenomenological phase map defined by the control parameters is considerably richer, and the interplay between impact, spreading, and retraction can lead to alternative bouncing behaviors or even complete suppression of bouncing, thereby transforming droplets into \emph{rollers}.

\begin{figure}[t!]
\centering
\includegraphics[width=1\textwidth]{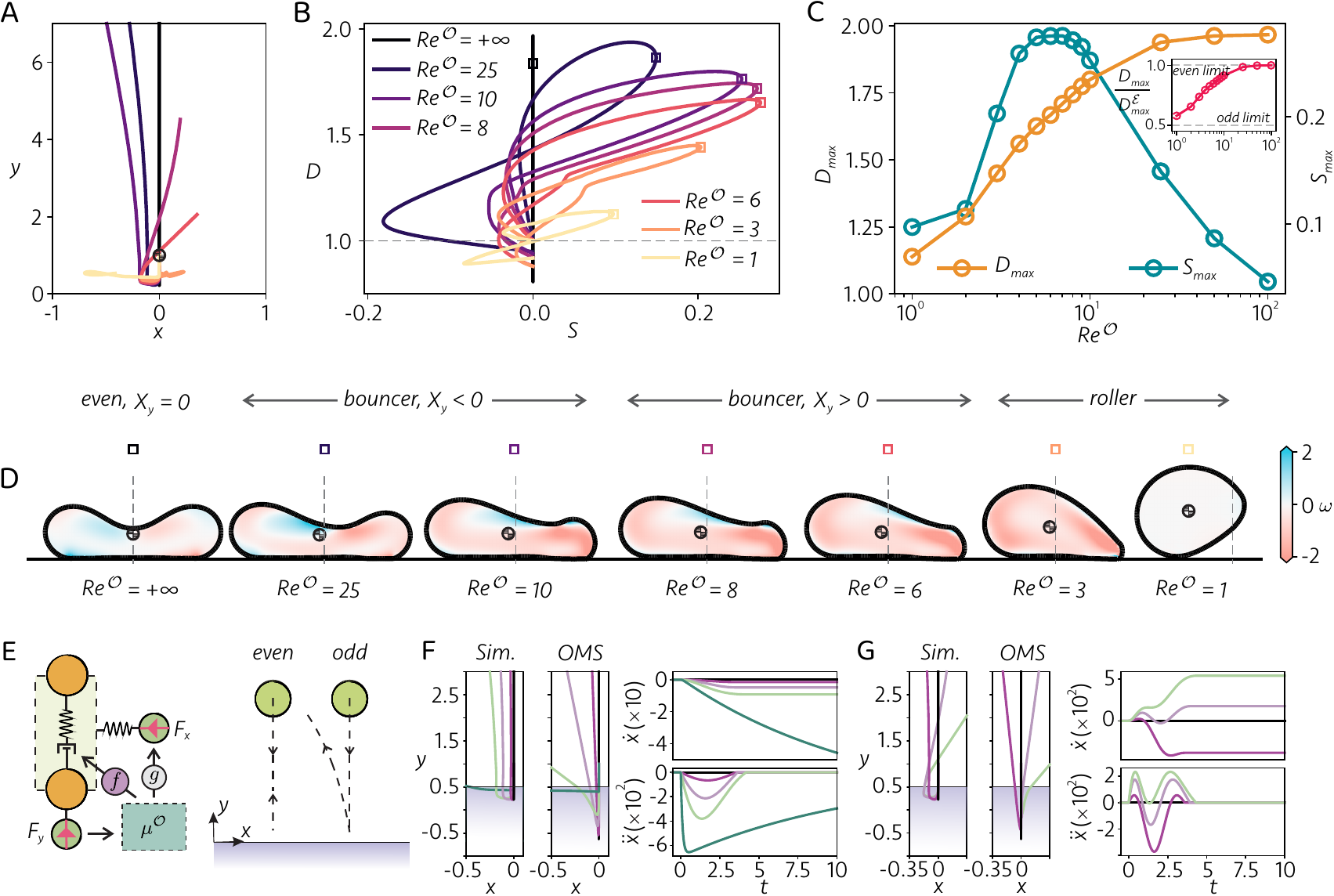}
\caption{\textbf{Controlling droplet dynamics with odd viscosity.} 
\textbf{A.} Depending on $Re^{\mathcal{O}}$, an odd droplet bounces leftwards, rightwards, or rolls on the surface, whereas an even droplet (black line) spreads, retracts, and bounces symmetrically. 
\textbf{B.} The spreading diameter and shape symmetry parameter illustrate the influence of $Re^{\mathcal{O}}$ on droplet dynamics. As the droplet spreads and retracts, its shape changes under the combined effects of inertial, capillary, and odd forces. Decreasing $Re^{\mathcal{O}}$ always leads to a smaller maximal diameter but a nonmonotonic maximal shape asymmetry. This is highlighted in panel \textbf{C}, where the maximal diameter varies between two limits: $D_{max} \rightarrow 1$ for $Re^{\mathcal{O}} \lesssim 1$ and $D_{max} \rightarrow D_{max}^{\mathcal{E}}$ for $Re^{\mathcal{O}} \gg 1$. Nonetheless, the maximal shape asymmetry $\mathcal{S}_{max}$ approaches zero for both large and small $Re^{\mathcal{O}}$, corresponding to the even and purely chiral limits, respectively. 
\textbf{D.} The shape of droplets with $\mathcal{S} = \mathcal{S}_{max}$ illustrates the deviation from even-symmetric cases. As $Re^{\mathcal{O}}$ decreases (and the effects of odd viscosity become more pronounced), the droplet becomes less symmetric and bounces to the left. With a further decrease in $Re^{\mathcal{O}}$, when odd viscosity dominates the flow, the shape asymmetry declines, causing the droplet to bounce to the right and eventually transform into a nearly circular roller. See \textbf{Supplementary Video II} for a dynamic visualization of all cases shown in this figure. $We=10$ for all results in this figure. \textbf{E.} An odd mass-spring (OMS) system provides a simple model of odd droplet impact: odd viscosity leads to left-right symmetry breaking and dissipation through flow modulation. \textbf{F.} Results of simulations are shown in panel \textbf{A} for $Re^{\mathcal{O}} = \{+\infty, \ 100, \ 25, \ 10, \ 1\}$. The simple OMS model captures the left bounce and the transition to overdamped, fully suppressed droplets; nonetheless, differences remain apparent. The droplet in the OMS model experiences horizontal acceleration ($\ddot{x}$) while interacting with the wall, leading to a horizontal velocity ($\dot{x}$) that remains constant after bouncing (here, constants are $c_1 = 9.3$, $c_2 = 0.63$, $k = 0.78$, $c_3=0$). \textbf{G.} Introducing a nonlinear function $g$ allows capturing both leftward and rightward bounces, as the total horizontal acceleration can change sign. Here the simulations are for $Re^{\mathcal{O}} = \{+\infty, \ 25, \ 8, \ 6\}$ and model constants are $c_1 = 6.04$, $c_2 = 0.63$, $k = 0.78$, $c_3 = 3$.
}
\label{fig2:changing_Re_odd}
\end{figure}
\begin{figure}[htbp]
\centering
\includegraphics[width=0.65\textwidth]{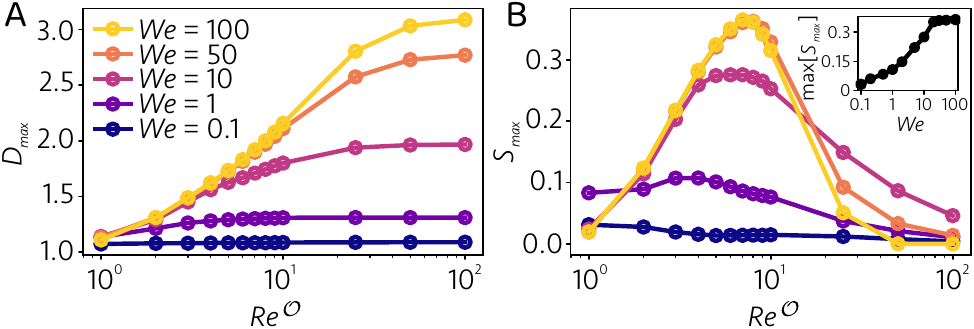}
\caption{
\textbf{Morphology dependence on surface tension and odd viscosity.} Variation in \textbf{A.}  maximal droplet spreading and \textbf{B.} maximal shape asymmetry for various values of $Re^{\mathcal{O}}$ and $We$. The inset in panel \textbf{B} displays the variation of the peak in $\mathcal{S}_{max}$ as a function of $We$.
}
\label{fig3:sweep_Re_We}
\end{figure}

Figure~\ref{fig2:changing_Re_odd} shows the effect of odd viscosity, varying $Re^{\mathcal{O}} \in [1, +\infty]$, while keeping $Re^\even = 50$ and $We = 10$ constant. Overall, three main dynamical categories emerge. In the limit of weak odd effects (\emph{i.e.}, $Re^{\mathcal{O}} \rightarrow \infty$), the droplet bounces symmetrically, as expected for a purely even droplet. Remarkably, the direction of left-right symmetry breaking in the trajectories of odd droplets (whether they bounce leftward or rightward) depends on the specific value of $Re^{\mathcal{O}}$. 
For large values of $Re^{\mathcal{O}}$ (corresponding to small odd viscosity), the droplet’s center of mass translates leftward during spreading and continues in that direction after bouncing. When $Re^{\mathcal{O}}$ is furthered reduced (\emph{i.e.} odd viscosity increases), the initial leftward translation during spreading persists; however, the retraction phase induces an opposing acceleration that ultimately drives the droplet rightward upon bounce.
We quantify the lateral preference of droplets using the parameter $X_y = \partial x/\partial y$ measured shortly after the bounce, which exhibits a transition from negative to positive values as $Re^{\mathcal{O}}$ varies.
A second remarkable phenomenon is the complete suppression of bouncing by odd viscosity. In cases of very high odd viscosity ($Re^{\mathcal{O}} \lesssim \mathcal{O}(1)$), the emergent odd force field results in a strong opposing flow field and effectively \say{lock} the droplets upon impact. Consequently, the chiral droplet rolls over the rigid surface with an internal rotational flow and purely horizontal translational velocity, much like a wheel (see Figure~\ref{figAppendix:visualize_fields_roll} in Appendix~\ref{appendix:velocity_field_distributions} for a demonstration of the flow fields). The morphology of these \emph{Rollers} remains nearly circular with surface perturbations and often initially moves leftward with a low speed.
%
The droplet dynamics described above can be interpreted through a simple odd mass-spring (OMS) model (see Fig.~\ref{fig2:changing_Re_odd}E). A conventional mass-spring system impacts and rebounds vertically from a wall~\cite{biance2006elasticity}. In contrast, the odd mass-spring model incorporates nonlinear coupling between compression and transverse motion due to odd viscosity. For $t < 0$, the droplet falls vertically at a constant speed $U$. For $0 < t < t_b$, where $t_b$ is the bouncing time, the droplet interacts with the wall ($y < 0.5$). During this period, horizontal forces are exerted, breaking the line symmetry of the trajectory. Additionally, odd viscosity modulates the flow field, leading to energy dissipation and suppression of bouncing. This interaction with the wall can be formulated as:
\begin{gather}
    m\,\ddot{y} + f(\mu^{\mathcal{O}})\,\dot{y} + k\,\left(y - 0.5\right) = 0
    \label{eq:odd_theory1}\\
    m\,\ddot{x} + g(\mu^{\mathcal{O}})\,y = 0
    \label{eq:odd_theory2}
\end{gather}
with initial conditions $x = \dot{x} = 0$, $y=0.5$, and $\dot{y} = -U$. Eqs. \ref{eq:odd_theory1} and \ref{eq:odd_theory2} represent a standard damped harmonic oscillator in the $y$-direction, coupled with a parity-breaking forcing in the $x$-direction. We assume for $t > t_b$, the droplet continues its trajectory with zero acceleration, \emph{i.e.}, $\ddot{x} = \ddot{y} = 0$. The choice of functions $f$ and $g$ determines the system’s dynamics. A simple choice is $f = c_1\,|\mu^{\mathcal{O}}|$ and $g = -c_2\,\mu^{\mathcal{O}}(1 - c_3\,y^2)$, where $c_i$ ($i = 1, 2, 3$) are constants.
For small vertical deformation, we have $g \approx -c_2\,\mu^{\mathcal{O}}$, and the mass accelerates horizontally to the left. If the vertical deformation is large, the horizontal force may switch sign, causing the mass to accelerate to the right. Additionally, if the magnitude of dissipation in the $y$-direction, due to flow modulation caused by odd viscosity, is sufficiently large, bouncing is completely suppressed.
In Fig.~\ref{fig2:changing_Re_odd}F and G, we compare the results of this simplified model with simulations. Despite its simplicity, the model captures the general trajectory dynamics. Nevertheless, notable differences arise, as it neglects even viscous dissipation and, more importantly, does not account for surface deformation or heterogeneous flow fields.


We further quantify the droplet's dynamic morphology using the parametric $D-\mathcal{S}$ curve shown in Fig.~\ref{fig2:changing_Re_odd}B.
Initially, the droplets exhibit a negative $\mathcal{S}$, as they skew to the left during the spreading. Later, during retraction, the skewness becomes positive, with maximum values of $\mathcal{S}$ (indicative of the most asymmetric shape) occurring in this phase (see the snapshots of these moments in Fig.~\ref{fig2:changing_Re_odd}D). 
Interestingly, the peak in $\mathcal{S}$ nearly coincides with the moment of maximum droplet diameter $D_{max}$, as the highly deformed droplet at that instant allows for strong asymmetry. Subsequently, surface tension retracts the droplet and promotes circularity, hence weakening these asymmetries ($[D,\mathcal{S}] \rightarrow [1,0]$ as $t\rightarrow \infty$). 
The values of $D_{max}$ and $S_{max}$ are of particular interest. When considering odd viscosity as a design parameter, the maximal spreading and maximal asymmetry appear as the two most important geometrical values. Fig.~\ref{fig2:changing_Re_odd}C shows how these values change with $Re^{\mathcal{O}}$. Notably, odd viscosity generates overall force fields that suppress the initial spreading of the droplet, always leading to smaller diameters than those of its even counterpart. The maximal droplet spreading has two distinct limits: for negligible odd viscosity ($Re^{\mathcal{O}} \rightarrow \infty$), the spreading reaches the even Newtonian values determined by the balance of surface energy, viscous dissipation, and kinetic energy~\cite{pasandideh1996capillary,sanjay2024unifying,naraigh2023analysis}. 
For highly odd droplets ($Re^{\mathcal{O}}\lesssim 1)$), the locking phenomena described above lead to the limit of $D_{max} \rightarrow 1$, \emph{i.e.}, the droplet remains circular.
The shape symmetry parameter $\mathcal{S}$, however, exhibits a non-monotonic trend. As odd viscosity increases to moderate values, the droplet becomes more asymmetric because the velocity field is significantly altered yet not fully converted into rotation. For more extreme odd viscosity values, nearly the entire velocity field transitions to solid-body rotation so that the droplet experiences little deformation and retains a nearly circular shape with minimal asymmetry (rollers).

The variations in $D$ and $\mathcal{S}$ also depend on the Weber number $We$, which here serves as a measure of surface tension strength (see Fig.~\ref{fig3:sweep_Re_We}). For small values of $We$, odd viscosity exerts only a minimal effect on these observables, as the dominant surface tension suppresses spreading and limits deformation, thereby preventing significant asymmetry.  As $We$ increases, inertial effects become more pronounced relative to capillarity, allowing the droplet to deform sufficiently for odd viscosity to influence its dynamics. Notably, $D_{max}$ always increases monotonically with $Re^\odd$ across all $We$ values, underscoring the role of odd viscosity in suppressing droplet spreading. A plateau is always reached for high $Re^\odd$, corresponding to the even limit. In contrast, the shape symmetry parameter $\mathcal{S}$ exhibits a nonmonotonic trend with odd viscosity for intermediate and high $We$ values. A peak in asymmetry is observed at intermediate $Re^{\mathcal{O}}$, where odd viscosity is sufficiently large to disrupt conventional symmetry but not so high as to nearly eliminate droplet deformation.  At large $We$ values, this maximum asymmetry reaches a plateau as surface tension effects disappear for $We\gg 1$ (see the inset in Fig.~\ref{fig3:sweep_Re_We}B).



A complete theoretical description (beyond the OMS model above) of the dependence of the three odd regimes observed above, namely, left bouncers ($X_y<0$), right bouncers ($X_y>0$), and rollers on the control parameters ($Re^{\mathcal{O}}$, We, and $Re^{\mathcal{E}}$) is not trivial,  particularly due to the highly nonlinear coupling among large interface deformations, inertia, and odd viscous effects and requires further investigations. 
In Appendices~\ref{appendix:velocity_field_distributions}, we further inspect the internal droplet vorticity field and its corresponding odd force field $\bm{F}^{\mathcal{O}}$ during the different phases of impact. Here, we proceed to perform a large-scale parametric study to identify the regime maps corresponding to these behaviors.


\section*{Summary \& Perspective:
Odd Viscosity to control interfacial dynamics and\\
design meta-fluids}

\begin{figure}[htbp]
\centering
\includegraphics[width=0.7\textwidth]{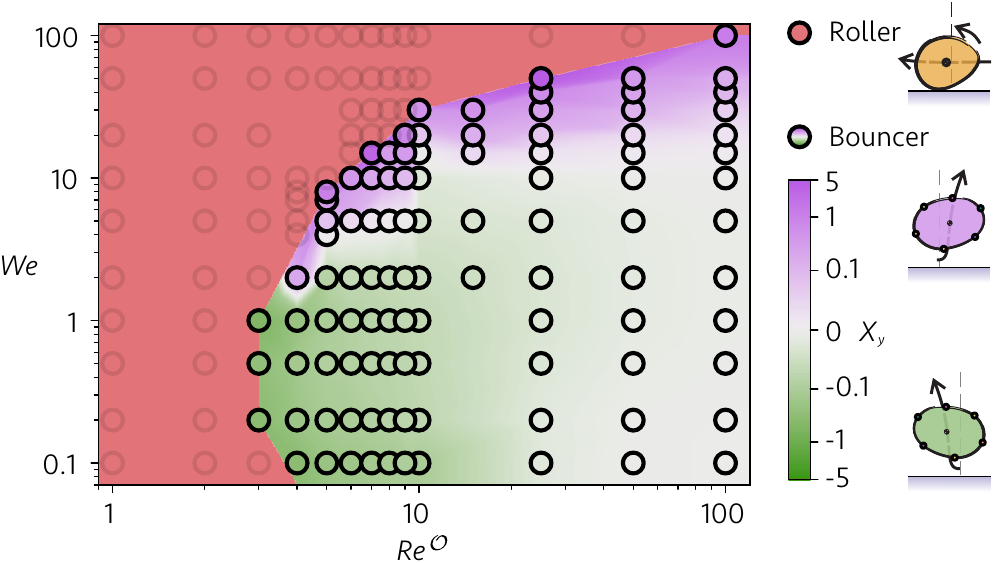}
\caption{
\textbf{Regime map of odd droplets.} Left and right bouncers, as well as rollers, are classified based on the balance among inertial, viscous, capillary, and odd forces. The left/right behavior of the bouncers is determined by the bouncing angle $X_y$, measured immediately after the droplet leaves the surface. 
}
\label{fig4:phasemap}
\end{figure}

The results presented above suggest that odd viscosity can be used as a tool to control the shape and dynamics of droplets. In Figure~\ref{fig4:phasemap}, we summarize the simulation results in a $We-Re^{\mathcal{O}}$ phase map. Four regimes are distinguishable: at low $Re^\odd$ values, odd viscosity dominates, locks the chiral droplets, suppresses bouncing, and, regardless of $We$ values, turns the droplets into rollers. At moderate $Re^\odd$ values, all regimes may be observed, depending on $We$. When $We$ is small, the odd force field during spreading dominates, causing the droplet to bounce leftward $X_y<0$; but, as $We$ increases, the force field emerged from a highly flattened droplet and its retraction becomes significant, and the droplet moves rightward. Surprisingly, at high $We$ values, large surface deformations result in pronounced odd viscosity effects and the dominance of chiral flows, leading to the reappearance of rollers. As $Re^{\mathcal{O}}$ increases further, all regimes persist, albeit more weakly, as even behavior is asymptotically recovered.

Our study employs direct numerical simulations of a continuum model to explore the highly nonlinear dynamics of droplets with odd viscosity and deformable interfaces under capillary stresses. The method, in principle, can be applied to any problem with deformable interfaces, and unlike models with geometrical constraints, typical of lubrication or shallow water theory, our approach provides access to solutions that remain out of reach for various analytical techniques. This capability not only enables detailed comparisons with experimental observations but also offers high-resolution insight into parameters such as strain rates and stress fields, quantities that are difficult to measure in laboratory settings. 

In a broader context, we show that introducing odd viscosity offers a rich phenomenology by drastically altering the dynamics of flows with deformable interfaces, as exemplified by the dynamics of droplets. Although odd viscosity is non-dissipative, it disrupts the conventional symmetric spreading observed in Newtonian fluids by modifying internal flow patterns and inducing asymmetric interface deformations, which in turn further alter the flow and the influence of odd viscosity. This nonlinear self-reinforcing loop eventually results in droplets self-control upon contact with a boundary. This control over droplet behavior, whether in the form of reduced maximal spreading or leftward/rightward bouncing or rolling motion, highlights the potential of odd viscous effects as a powerful tool for controlling multiphase flows. This ability may have promising industrial applications in the future. 
Consider, for example, inkjet printing, where suppressing unwanted bouncing and precisely directing droplets is of great interest. However, current droplet control methods rely on external actuation, such as temperature and electric fields, or on modifying boundary conditions like using asymmetrically structured surfaces~\cite{jalaal2014controlled,jalaal2018gel,ahmed2018maximum,lohse2022fundamental,yang2022droplet,jezervsek2023control,zhao2023regulating,ramesh2025frozen}. Controlling droplets, and in general interfacial fluids, with odd viscosity may overcome the limitations associated with these techniques by enabling the fluid to self-regulate based on the governing physical parameters. This, nonetheless, requires the design of chiral fluids in driven or active systems and their subsequent implementation in engineering contexts, a challenge that still demands further research and development.
Looking even further into the future, the development of meta-fluidic materials that leverage activity and odd properties to interact with complex environments presents another exciting frontier. To this end, it is possible to complement the recent advances in adaptive active solids~\cite{Veenstra2025Adaptive}, while exploring opportunities beyond the constraints of elasticity.

There are several immediate opportunities to extend the present work. First, developing minimal models of interfacial fluids that incorporate odd viscosity and large deformations could provide further theoretical insights into these complex behaviors. 
The odd mass-spring model presented here, despite capturing key features of odd droplet dynamics, remains limited in several aspects, such as accurately describing trajectories, revealing shape asymmetries,  and finding the boundaries of transition to different regimes. Addressing these limitations represents a promising direction for future research. 
Second, integrating odd viscosity with other mechanical properties, particularly elasticity~\cite{banerjee2021active,lier2022passive,Veenstra2025Adaptive}, may not only lead to tools for optimizing the design of functional metamaterials but also deepen our understanding of their nonlinear behavior. In fact, the intriguing similarities and differences between odd viscous fluids and odd elastic bodies with deformable interfaces~\cite{Veenstra2025Adaptive} merit further investigation. Third, exploring alternative forms of odd viscous terms arising from microscopic constituents could enable tailored solutions for specific fluidic applications~\cite{souslov2020anisotropic}. Finally, the implications of our results may extend beyond classical fluid mechanics and present modeling opportunities in other condensed matter problems, such as electronic fluids, where interactions with boundaries play a crucial role in system dynamics~\cite{souslov2019topological,fritz2024hydrodynamic,tauber2020anomalous}.

\section*{Acknowledgements}
We thank Tom Appleford, Pedro Cosme, Vatsal Sanjay, Ruben Lier, Jonas Veenstra, and Corentin Coulais for valuable discussions.
M.J. acknowledges support
from the ERC grant no. ”2023-StG-101117025, FluMAB.” This publication is part of the Vidi project Living
Levers with file number 21239, financed by the Dutch Research Council (NWO). We acknowledge the computational support given by FCT-UNESP through the usage of their cluster LSNCS for these simulations.
\pagebreak
\section*{Appendices}
\subsection{Equations \& Boundary conditions}
\label{appendix:Equations}

The flow inside and outside the droplet is modeled by the two-phase incompressible Navier-Stokes equations, given by:
\begin{align}
& \bm{\nabla} \cdot \bm{u} = 0, \label{eq:navier1}\\
& \rho\left( \frac{\partial \bm{u}}{\partial t} + ( \bm{u} \cdot \bm{\nabla} ) \, \bm{u} \right)	= - \bm{\nabla} p  + \bm{\nabla} \cdot \bm{\tau}^\even + \bm{\nabla} \cdot \bm{\tau}^\odd + \bm{F}^{\mathcal{S}}, \label{eq:navier0}
\end{align}
where $\bm{u}$ and $p$ are the velocity and pressure fields, respectively, and $\rho$ is the fluid mass density. In our numerical method, the surface tension force is defined as a body force $\bm{F}^{\mathcal{S}} = \sigma \kappa \delta_s \bm{n}$, where $\kappa$ is the curvature of the interface, $\sigma$ the constant surface tension coefficient, $\bm{n}$ is the unit vector normal to the interface, and $\delta_s$ is the Dirac delta function centered on the interface~\cite{Tryggvason2011-book}. The deviatoric stress tensor of the fluid is $\bm{\tau} =  \bm{\tau}^\even + \bm{\tau}^\odd$, where the even and odd components of stress are given by
\begin{align}
\bm{\tau}^\even = \mu^\even\dot{\bm{\gamma}}^\even & = \mu^\even \left[\begin{array}{cc} 2\,\del{u}{x} & \del{u}{y} + \del{v}{x} \\ \del{u}{y} + \del{v}{x}& 2\,\del{v}{y} \end{array}\right], \label{eq:tau_even}\\
\bm{\tau}^\odd = \mu^\odd\dot{\bm{\gamma}}^\odd & = \mu^\odd \left[\begin{array}{cc} -\left( \del{u}{y} + \del{v}{x} \right) & \del{u}{x} - \del{v}{y} \\ \del{u}{x} - \del{v}{y} & \del{u}{y} + \del{v}{x} \end{array}\right] \label{eq:tau_odd},
\end{align}
and $\mu^\even$ and $\mu^\odd$ are the even and odd viscosity coefficients, respectively. 

On the rigid surface, no-slip and non-wetting boundary conditions are applied. Outflow conditions are applied for the other three boundaries of the domain.  The interface between the two fluids is tracked by the Volume of Fluid (VOF) scheme \cite{Hirt1981}. In this method, a tracer $c \in (0, 1)$ stores the fraction of droplet fluid contained in each cell of the mesh. The properties $\rho$, $\mu^\even$ and $\mu^\odd$ in Eqs. \eqref{eq:navier0}-\eqref{eq:tau_odd} are then interpolated locally as
\begin{align}
\rho(c) = & \  c \ \rho_d + (1 - c)\,\rho_a, \label{eq:vof_density_viscosity_1}\\
\mu^\even(c) = & \ c \  \mu_d^\even + (1 - c)\,\mu_a^\even, \label{eq:vof_density_viscosity_2}\\
\mu^\odd(c) = & \ c \,  \mu_d^\odd \label{eq:vof_density_viscosity_3},
\end{align}
where the subscripts $d$ and $a$ indicate, respectively, the properties of the droplet ($\Omega_1$) and the outer fluid ($\Omega_2$). 

We non-dimensionlize Eqs.\eqref{eq:navier1} and \eqref{eq:navier0} by defining rescaled variables as
\begin{equation}
    \textbf{x} = D_0\, \bar{\textbf{x}}, \hspace{10pt} t = \frac{D_0}{U}\bar{t}, \hspace{10pt} \bm{u} = U\bar{\bm{u}}, \hspace{10pt} p =  \rho_d\,U^2\bar{p}, \hspace{10pt} \kappa =  \frac{1}{D_0}\bar{\kappa}, \hspace{10pt} \delta_s =  \frac{1}{D_0}\bar{\delta_s}.
\label{eq:rescalings}
\end{equation} 

The dimensionless version of Eqs. \eqref{eq:navier1}-\eqref{eq:navier0} will then become
\begin{align}
&\bm{\nabla} \cdot \bar{\bm{u}} = 0, \label{eq:nondim_navier1}\\
& \bar{\rho}\left( \frac{\partial \bar{\Ubold}}{\partial t} + (\bar{\bm{u}} \cdot \bm{\nabla})  \bar{\bm{u}}  \right) 	= - \bm{\nabla} p  + \bar{\mu}^\even\frac{1}{Re^\even} \, \bm{\nabla} \cdot  \bar{\dot{\bm{\gamma}}}^\even + \bar{\mu}^\odd\frac{1}{Re^\odd}\, \bm{\nabla} \cdot  \bar{\dot{\bm{\gamma}}}^\odd + \frac{1}{We}\bar{\kappa}\bar{\delta}_s\bm{n}, \label{eq:nondim_navier0}\\
\end{align}
where the auxiliary variables $\bar{\rho}, \bar{\mu}^\even, \bar{\mu}^\odd$ are the dimensionless versions of the VOF interpolating functions in Eqs. \eqref{eq:vof_density_viscosity_1}-\eqref{eq:vof_density_viscosity_3}, given by
\begin{align}
    \bar{\rho}(c) = & \  c + (1 - c)\,\rho_a/\rho_d, \label{eq:nondim_vof_density_viscosity_1}\\
    \bar{\mu}^\even(c) = & \ c + (1 - c)\,\mu_a^\even/\mu_d^\even, \label{eq:nondim_vof_density_viscosity_2}\\
    \bar{\mu}^\odd(c) = & \ c. \label{eq:nondim_vof_density_viscosity_3}
\end{align}

Throughout this paper, we take the properties of the outer fluid as $\rho_a = 10^{-2}\rho_d$ and $\mu_a^\even = 10^{-2}\mu_d^\even$, such that our system is determined only by three parameters: the even Reynolds, odd Reynolds, and Weber numbers. These numbers are defined, respectively, by
\begin{equation}
    Re^\even = \frac{\rho_d \,U D_0}{\mu^\even}, \hspace{20pt} Re^\odd = \frac{\rho_d \, U D_0}{\mu^\odd}, \hspace{20pt} We = \frac{\rho_d \, U^2D_0}{\sigma}.
\label{eq:dimensionless_groups}
\end{equation}

In all other sections of this paper, only the nondimensional variables in \eqref{eq:rescalings} are used, and for convenience, we omit the upper bars.

\subsection{Numerical Implementation}
\label{appendix:numerical_implementation}

We used the open source free software language Basilisk C to solve the equations of motion \eqref{eq:nondim_navier1}-\eqref{eq:nondim_navier0}.
An overview of the numerical procedure will be given in this section, while detailed descriptions of Basilisk C can be found in \cite{Popinet2013-Basilisk}. The numerical methods used have been tested for various Newtonian and non-Newtonian fluids with deformable interfaces, including many droplet dynamics scenarios~\cite{popinet2009accurate,jalaal2012fragmentation,cimpeanu2018early,jalaal2021spreading,zhang2022impact,francca2024elasto,sanjay2025role}.

\begin{figure}[htbp]
\centering
\includegraphics[width=1\textwidth]{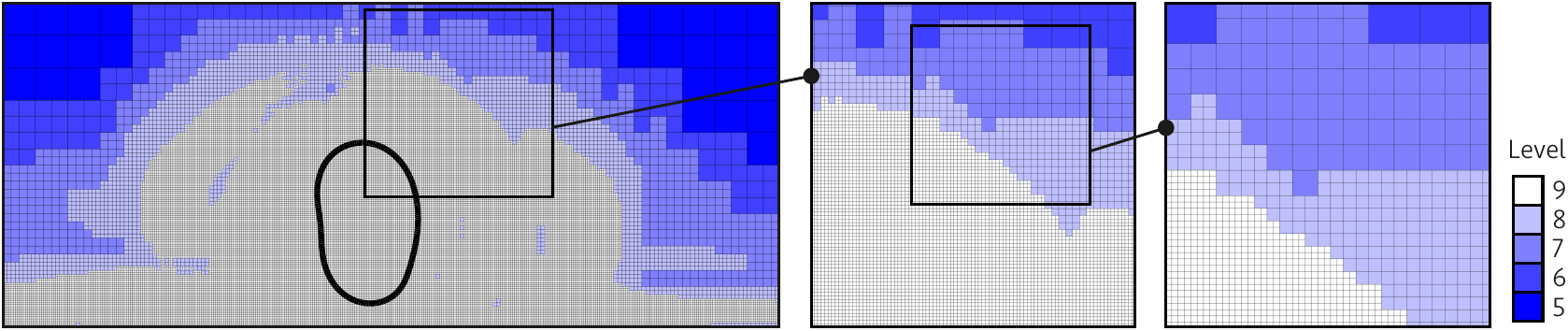}
\caption{\textbf{Example of mesh with adaptive refinement.} The mesh is based on a quadtree structure and the level of refinement is higher closer to the droplet, adapting over time as the droplet deforms and moves.}
\label{figAppendix:mesh}
\end{figure}

The simulation is setup by creating a square domain $[-4, 4] \times [0, 8]$ where a droplet of diameter $1$ is placed with its center at $[0, 1]$. The velocity within the droplet region ($\Omega_1$) is initialized as ${\bm{u}}_0 = [0, -1]$. To accurately resolve the flow structure inside the droplet and its shape, we apply increased refinement levels for the liquid phase and around it. The mesh is dynamically adapted as the droplet deforms over time. The additional refinement follows the interface and areas around it where the flow needs to be accurately resolved~\cite{Popinet2015}. For example, the mesh corresponding to the simulation from Fig. \ref{fig1:phenomenology} is shown in Fig. \ref{figAppendix:mesh} for a specific time instant. This figure illustrates how the mesh is created following a quadtree structure with a maximum level of refinement 9. Considering the total size of the square domain $(8)$, our smallest cells have size $\Delta_x = 8/2^9 = 0.015625$, such that the droplet diameter is discretized by $64$ cells. 

As mentioned in section \ref{appendix:Equations}, the droplet interface is tracked through a VOF tracer field $c$. This field is advected over time by solving the equation
\begin{equation}
    \del{c}{t} + \bm{\nabla} \cdot (c \, \bm{u}) = 0.
\label{eq:vof_advection}
\end{equation}

The standard Navier-Stokes solver in the Basilisk C language solves the governing equations using a projection method and a multilevel Poisson solver to treat the even viscosity term $\mu^\even \, \bm{\nabla} \cdot \bm{\tau}^\even$ (see \cite{popinet2009accurate,Popinet2015} for more details). In our implementation, we modify this standard code in order to add the new terms representing the odd viscous force $\mu^\odd \, \bm{\nabla} \cdot \bm{\tau}^\odd$. 


\subsubsection{Odd Poiseuille flow within a channel}
We validate our implementation with a simple single-phase scenario that allows for an analytical solution. A two-dimensional channel is created with half-width $H$ and length $L$. The parabolic Poiseuille velocity profile is imposed on the left side as a boundary condition. We nondimensionalize our equations by rescaling variables in the same way mentioned in Eqs. \eqref{eq:rescalings}, but using $H$ as length scale and the maximum velocity of the Poiseuille profile $U_{max}$ as velocity scale. In this manner, we obtain the same nondimensional equations \eqref{eq:nondim_navier1}-\eqref{eq:nondim_navier0}, but for a single fluid $(\bar{\rho} = \bar{\mu}^\even = \bar{\mu}^\odd = 1)$ and with no surface tension $(We \rightarrow +\infty)$. The solution will then depend only on the two Reynolds numbers $Re^\even$ and $Re^\odd$ and the boundary conditions. 

Note that we do not impose the pressure gradient as a condition, such that this gradient will be calculated by the solver from the imposed velocity field. It is, however, necessary to impose at least one boundary condition for the pressure to guarantee a unique solution. With this in mind, we impose $p=0$ at the right wall. While this condition is not consistent at the boundary with the analytical solution presented below, we will show that further from this boundary, the solver can still recover the correct analytical solution expected for a fully developed Poiseuille profile.

In this scenario, Eqs. \eqref{eq:nondim_navier1}-\eqref{eq:nondim_navier0} can be easily solved analytically to obtain the pressure and its gradient within the domain, which are given by
\begin{align}
    p(x, y) & = \frac{2}{Re^\even}\left(\frac{L}{H} - x \right) - \frac{2}{Re^\odd}y, \label{eq:poiseuille_pressure} \\
    \nabla p & =  \left[- \frac{2}{Re^\even}, \ \frac{2}{Re^\odd}\right]. \label{eq:poiseuille_pressure_grad}
\end{align}
Note that, for $Re^\odd \rightarrow \infty$, we recover the classical pressure from a Poiseuille flow, which presents simply a constant gradient in the $x$ direction. With the presence of odd viscosity, a constant gradient also appears in the $y$ direction. This classical solution has also been shown in \cite{Fruchart2023} and was similarly used as numerical validation in \cite{aggarwal2023thermocapillary}. 

\begin{figure}[htbp]
\centering
\includegraphics[width=0.8\textwidth]{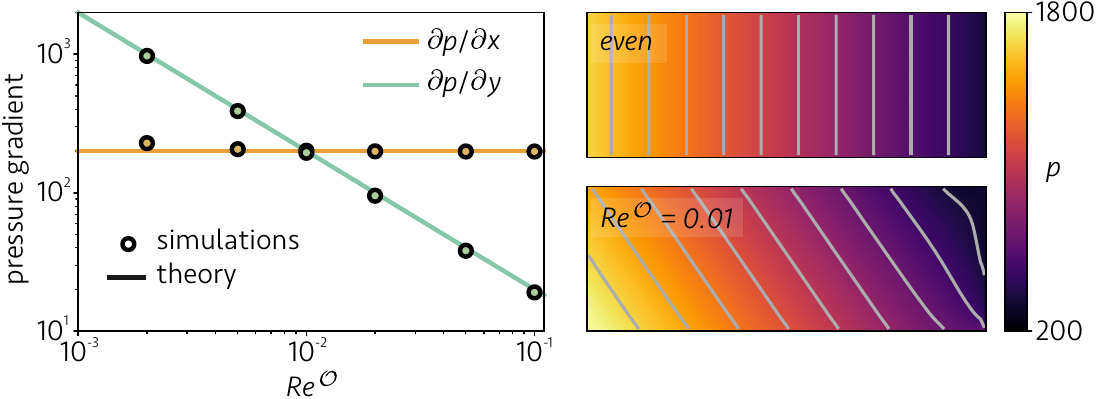}
\caption{\textbf{Pressure field and its gradients in a Poiseuille flow for a fluid with different odd viscosity.} \textbf{A.} Comparison between numerical (circles) and analytical (solid lines) values for the pressure gradient as a function of $Re^\odd$. \textbf{B.} The full pressure field in the channel is shown for an even and an odd fluid with $Re^\odd = 0.01$. The isolines of the pressure field are shown in gray.}
\label{figAppendix:poiseuille}
\end{figure}

Figure \ref{figAppendix:poiseuille}B shows the pressure field within the computational domain for an even and an odd fluid. In all cases, the even Reynolds number is kept fixed as $Re^\even = 0.01$, so that no inertial effects are expected. From these fields, we can see that odd viscosity introduces a pressure gradient in the $y$ direction as it rotates the pressure isolines from vertical to diagonal. We quantified the $x$ and $y$ components of this pressure gradient by taking a horizontal cross-section at the center $y = 0$ and a vertical at $x = L/(2H)$. This gradient is plotted for $6$ values of $Re^\odd$ in Fig. \ref{figAppendix:poiseuille}A, where we also show the analytical solution from Eq. \eqref{eq:poiseuille_pressure_grad}. We see that our solution quantitatively agrees with the expected values.

\subsection{Odd force fields during spreading and retraction}
\label{appendix:odd_force_distributions}
\begin{figure}[htbp!]
\centering
\includegraphics[width=1\textwidth]{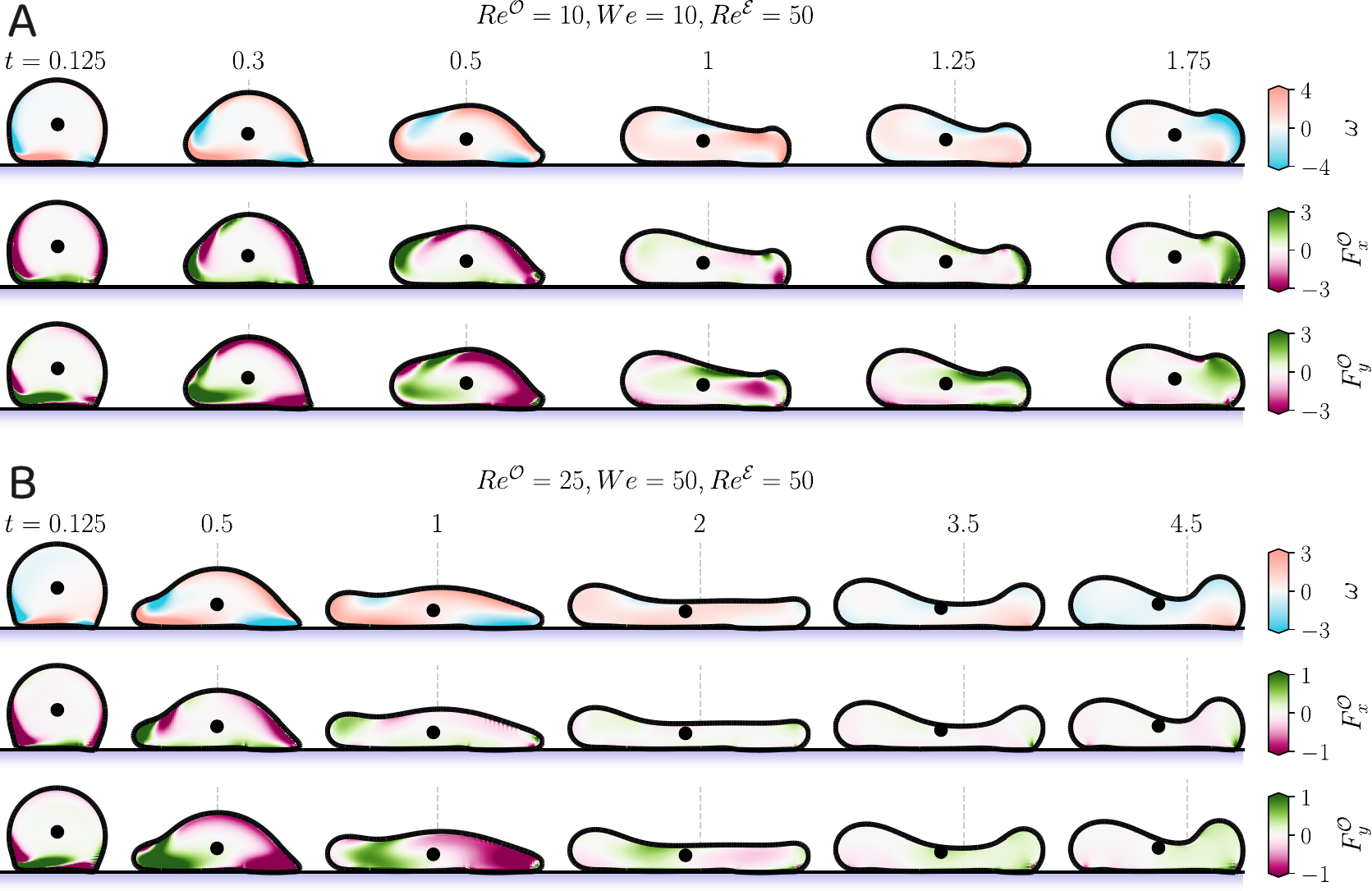}
\caption{\textbf{Vorticity and odd force fields during spreading and retraction.} \textbf{A.} An example of a left bouncer ($Re^\even = 50$, $Re^\odd = 10$, $We = 10$). \textbf{B.} An example of a right bouncer ($Re^\even = 50$, $Re^\odd = 25$, $We = 50$).}
\label{figAppendix:visualize_odd_forces}
\end{figure}

To elucidate the time-dependent, heterogeneous odd force field during the impact, spreading, and retraction phases, we present in Fig.~\ref{figAppendix:visualize_odd_forces} the fields for vorticity and both components of $\bm{F}^\odd$ at various time instants. Panel \textbf{A} displays the same example as in Fig.~\ref{fig1:phenomenology} of the main text. We observe that near the moment of impact, the horizontal odd force $F^{\mathcal{O}}_x$ exhibits localized positive regions close to the solid surface due to steep vorticity gradients, while it remains negative throughout the bulk of the droplet. This explains the characteristic leftward translation during impact. As the droplet spreads and deforms, positive values of $F^{\mathcal{O}}_x$ begin to emerge within the bulk and along the upper surface. This shift leads to a gradual rightward acceleration that persists during the retraction. In this specific case, however, the spreading and retraction periods are too short for the rightward acceleration to completely counteract the initial leftward velocity, and the droplet ultimately bounces while still translating to the left.

Panel \textbf{B} of Fig.~\ref{figAppendix:visualize_odd_forces} presents a second case in which the droplet bounces and translates to the right. Near the moment of impact, the behavior is almost identical to that in panel \textbf{A}. This brief timescale prevents significant viscous (both odd and even) and capillary effects from developing, so the droplet initially exhibits its characteristic odd force field with a predominantly negative  $F^{\mathcal{O}}_x$ component. However, due to higher values of $We$ and $Re^\odd$, the droplet undergoes prolonged spreading and retraction phases. During these extended stages, the emergence of positive $F^{\mathcal{O}}_x$ values eventually reverses the droplet's translation, leading to rightward motion at the moment of bouncing.

\pagebreak
\subsection{Velocity field distributions for characteristic scenarios}
\label{appendix:velocity_field_distributions}

\begin{figure}[htbp]
\centering
\includegraphics[width=1\textwidth]{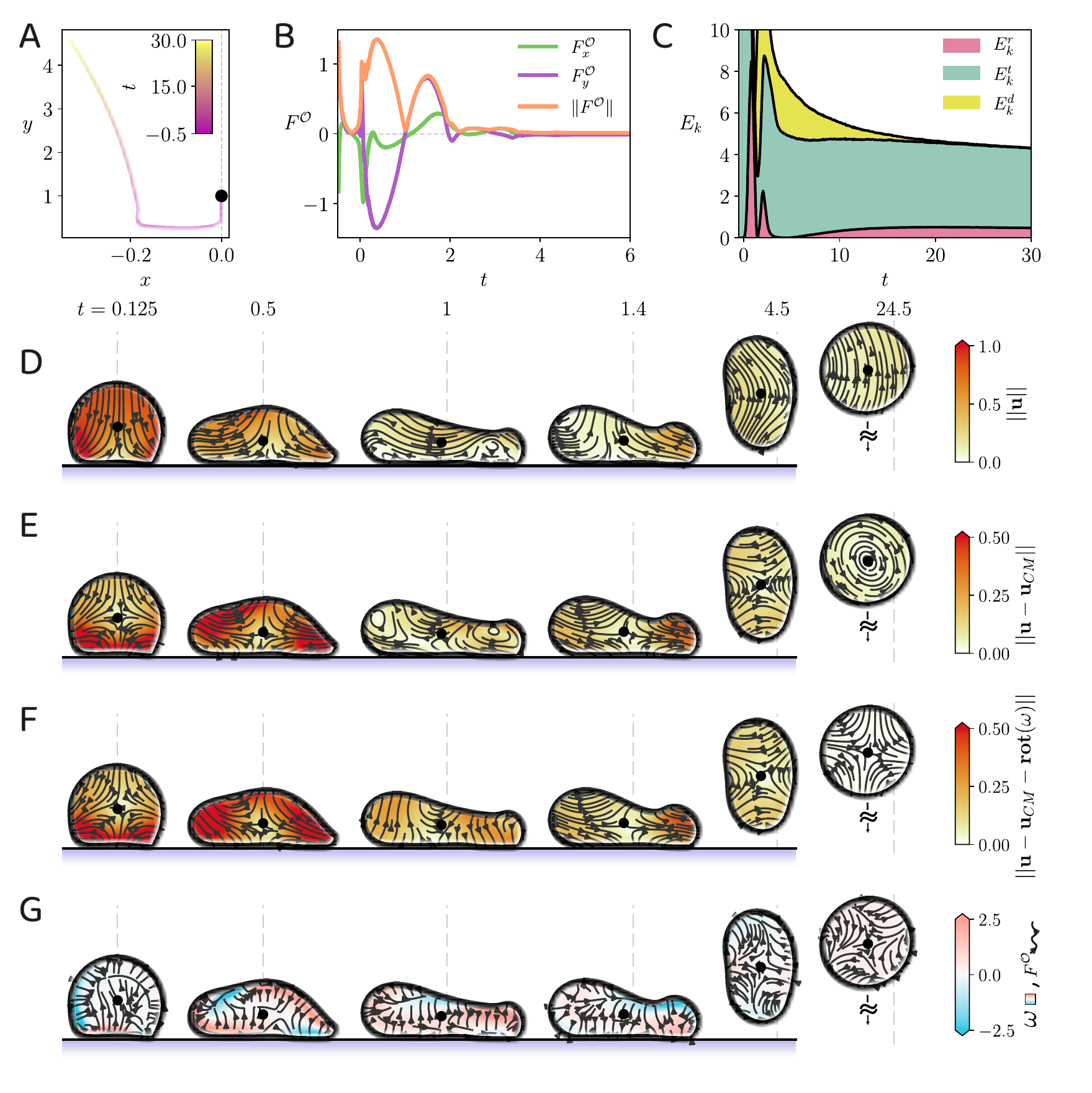}
\caption{\textbf{Dynamics of a left-bouncing odd droplet ($Re^\even = 50$, $Re^\odd = 10$, $We = 10$).} \textbf{A.} Center-of-mass trajectory, marked by time. \textbf{B.} Components of the average odd force over time. \textbf{C.} Kinetic energy budget over time, where superscripts $r$, $t$, and $d$ denote rotation, translation, and deformation, respectively. At late times, nearly the entire velocity field consists of translation and solid-body rotation. \textbf{D.} (Lab-frame) velocity fields at selected time instants. \textbf{E.} Velocity fields with the pure translation component removed, revealing the late-time solid-body rotation induced by odd viscosity. \textbf{F.} Deformation component of the velocity field. \textbf{G.} Vorticity fields (background colors) and the odd force (streamlines). The odd force follows the negative gradient of the vorticity ($F^\odd = -\bm{\nabla} \omega$). See \textbf{Supplementary Video III} for a dynamic visualization of this simulation.}
\label{figAppendix:visualize_fields_bounce_left}
\end{figure}

A closer inspection of the dynamics and internal flow fields of the odd droplet is provided for four characteristic scenarios: (i) leftward bouncing, (ii) rightward bouncing, (iii) surface rolling, and (iv) an extreme case of rolling, where the droplet slides and comes to a stop.
In Fig.\ref{figAppendix:visualize_fields_bounce_left}, we inspect the velocity field for an impact resulting in a droplet that bounces to the left. This case corresponds to the parameters $Re^\even=50$, $Re^\odd=10$, and $We=10$ (same as in Fig.~\ref{fig1:phenomenology} and Fig.~\ref{figAppendix:visualize_odd_forces}A). Immediately after impact, the highly inertial flow creates a strong localized shear that generates an odd force, pushing the droplet to the left. This is evidenced by the sharp negative peak in $F^\odd_x$ shown in Fig.\ref{figAppendix:visualize_fields_bounce_left}A, which consistently appears shortly after impact.
As the droplet spreads, the peak in $F^\odd_x$ softens as the velocity field changes and its magnitude decreases. The complex interplay of inertial spreading, capillary retraction, and viscous dissipation eventually reverses the sign of the average $F^\odd_x$, leading to an acceleration of the droplet to the right that reduces the initial leftward velocity obtained during impact. However, this acceleration is not strong enough to completely reverse the droplet’s motion, so it moves left after bouncing.
As the droplet translates upward, it experiences shape oscillations driven by surface tension acting on its non-spherical form. Simultaneously, a small drag force is present due to the low-density outer fluid. The velocity field generated by these oscillations and the drag produces a small odd force that, over time, modifies the droplet’s internal velocity field. At late time, as shown in Panel \textbf{D}, the velocity field of the bouncing droplet appears to be almost pure upward translation. However, after subtracting the center-of-mass velocity, $\bm{u}_{CM}$, the residual velocity field reveals an almost perfect solid-body rotation, indicating that the droplet is both translating and rotating.
By measuring the average vorticity $\omega_{\text{avg}}$ within the droplet and subtracting the solid-body rotation field, $\textbf{rot}({\omega}_{avg}) = -\frac{1}{2}{\omega}_{avg}[- ({y} - {y}_{CM}), \ {x} - {x}_{CM}]$, we obtain the final residual velocity field in the droplet, which still has the potential to induce deformation. Note that the translational and solid-body rotational components do not trigger odd viscosity; thus, the residual velocity field from Panel \textbf{F} is the only part capable of generating odd forces.

At very late times, nearly the entire velocity field is converted into either translation or pure rotation, resulting in a negligibly small residual velocity field and eliminating any significant odd viscosity effects. To quantify the three components of the velocity field, Panel \textbf{C} plots the total kinetic energy within the droplet, separated by the energy contributions from each component. As expected, at late times, all kinetic energy is converted into translational and rotational energy. We note that, in a Newtonian (even) scenario, no rotation would occur since the droplet would maintain perfect left-right symmetry with zero average vorticity. Therefore, the presence of a rotational component in the kinetic energy is a direct consequence of odd viscosity.

\begin{figure}[htbp]
\centering
\includegraphics[width=1\textwidth]{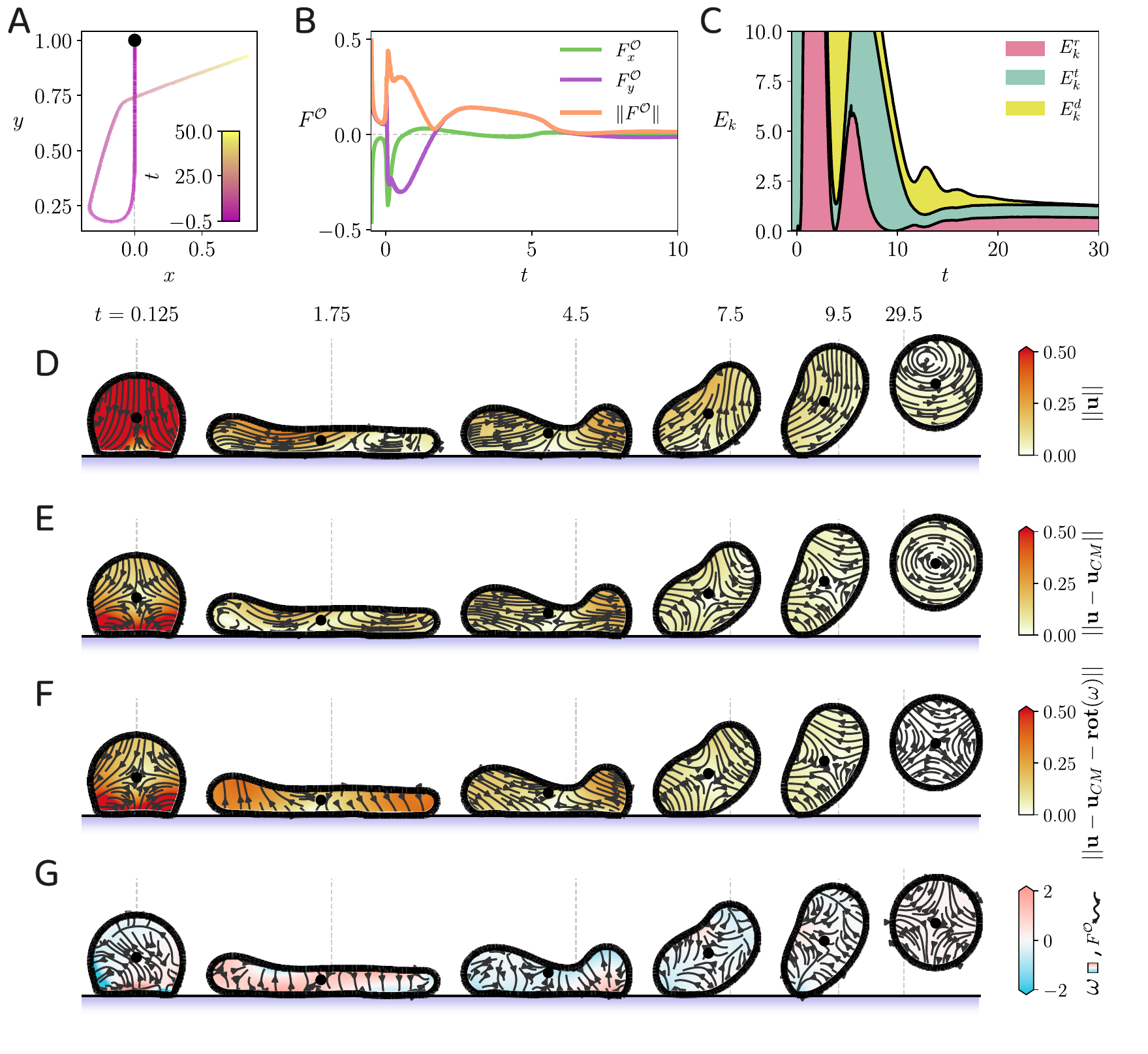}
\caption{\textbf{Same as above for dynamics of a right-bouncing odd droplet ($Re^\even = 50$, $Re^\odd = 25$, $We = 50$).} See \textbf{Supplementary Video IV} for a dynamic visualization of this simulation.}
\label{figAppendix:visualize_fields_bounce_right}
\end{figure}

Figure~\ref{figAppendix:visualize_fields_bounce_right} shows the velocity fields for an impact with $Re^\even=50$, $Re^\odd=25$, and $We=50$, resulting in a droplet that bounces to the right. Immediately after impact, a negative peak in the average $F^\odd_x$ is observed, as in the previous case (Fig.\ref{figAppendix:visualize_fields_bounce_left}). The key difference is the larger maximum spreading reached by this droplet, which leads to extended spreading and retraction phases. During these longer phases, the positive $F^\odd_x$ generated by the spreading and retraction is sufficient to accelerate the droplet to the right, overcoming the initial leftward velocity emerged at impact. Consequently, the droplet's horizontal translation is fully reversed to the right before it completely bounces off the surface. After bouncing, the subsequent dynamics are similar to those described for the case in Fig.~\ref{figAppendix:visualize_fields_bounce_left}.

\begin{figure}[htbp]
\centering
\includegraphics[width=1\textwidth]{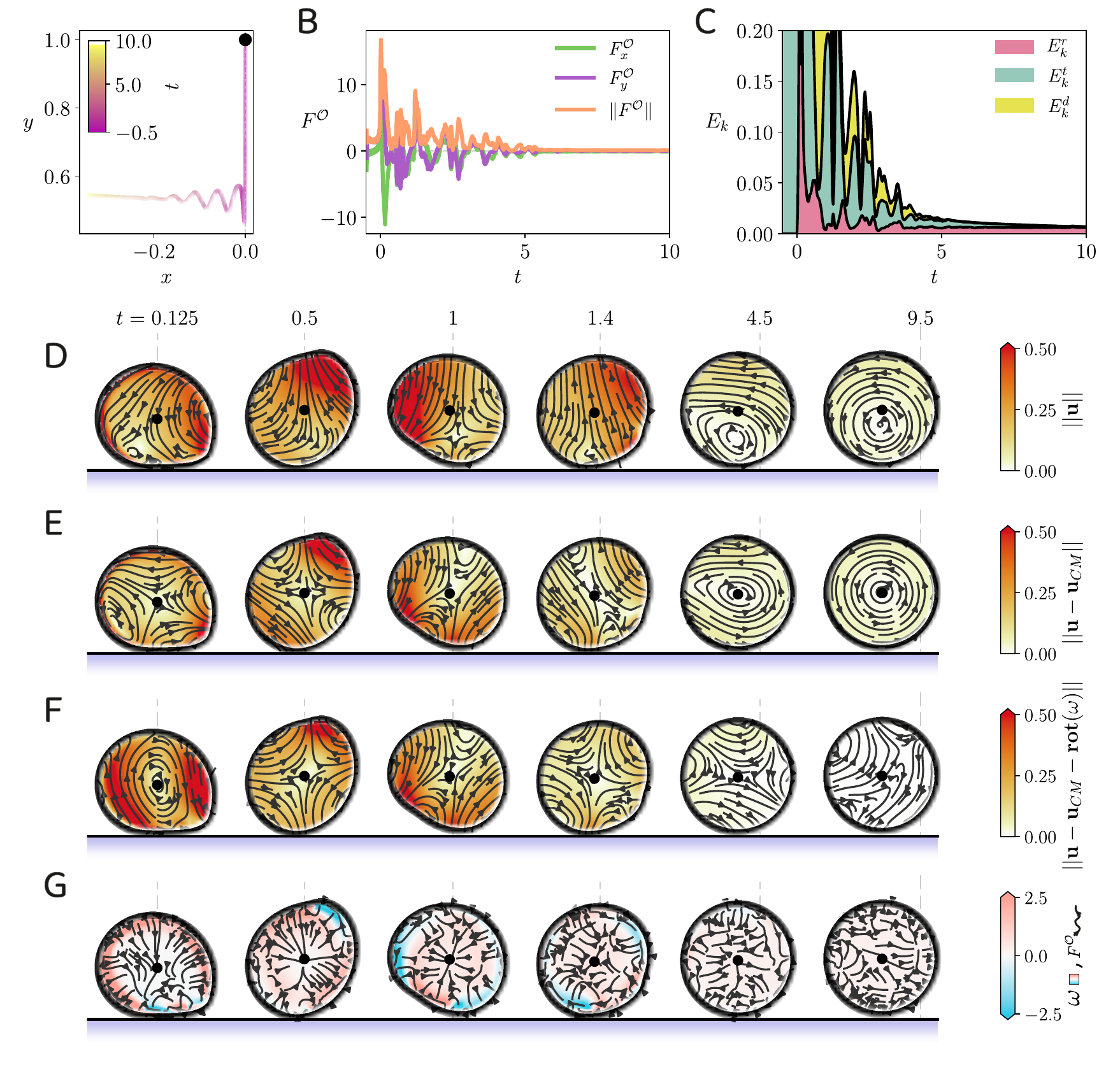}
\caption{\textbf{Same as above for a roller with high odd viscosity ($Re^\even = 50$, $Re^\odd = 1$, $We = 0.2$).} See \textbf{Supplementary Video V} for a dynamic visualization of this simulation.}
\label{figAppendix:visualize_fields_roll}
\end{figure}

Figure~\ref{figAppendix:visualize_fields_roll} shows the velocity fields for an impact resulting in a droplet that rolls over the surface. This case corresponds to $Re^\even=50$, $Re^\odd=1$, and $We=0.2$, representing a case with very high odd viscosity. The strong surface tension at this low Weber number, combined with the suppression of spreading due to high odd viscosity, results in minimal droplet deformation after impact. Because the deformation is so small, the droplet is unable to retract and bounce. Nonetheless, the internal flows during impact activate the odd viscosity, generating a rotational component in the velocity field, as revealed by the kinetic energy decomposition in Panel \textbf{C}. Over time, almost the entire velocity field is converted into solid-body rotation, so that the droplet simply rotates around its own axis on the rigid surface. Due to the superhydrophobic, non-wetting properties of the surface, the droplet’s interaction with it is weak, and this rotation barely produces any translation. If a friction mechanism were present between the droplet and the surface, one could expect the droplet to behave like a moving wheel.

\begin{figure}[htbp]
\centering
\includegraphics[width=1\textwidth]{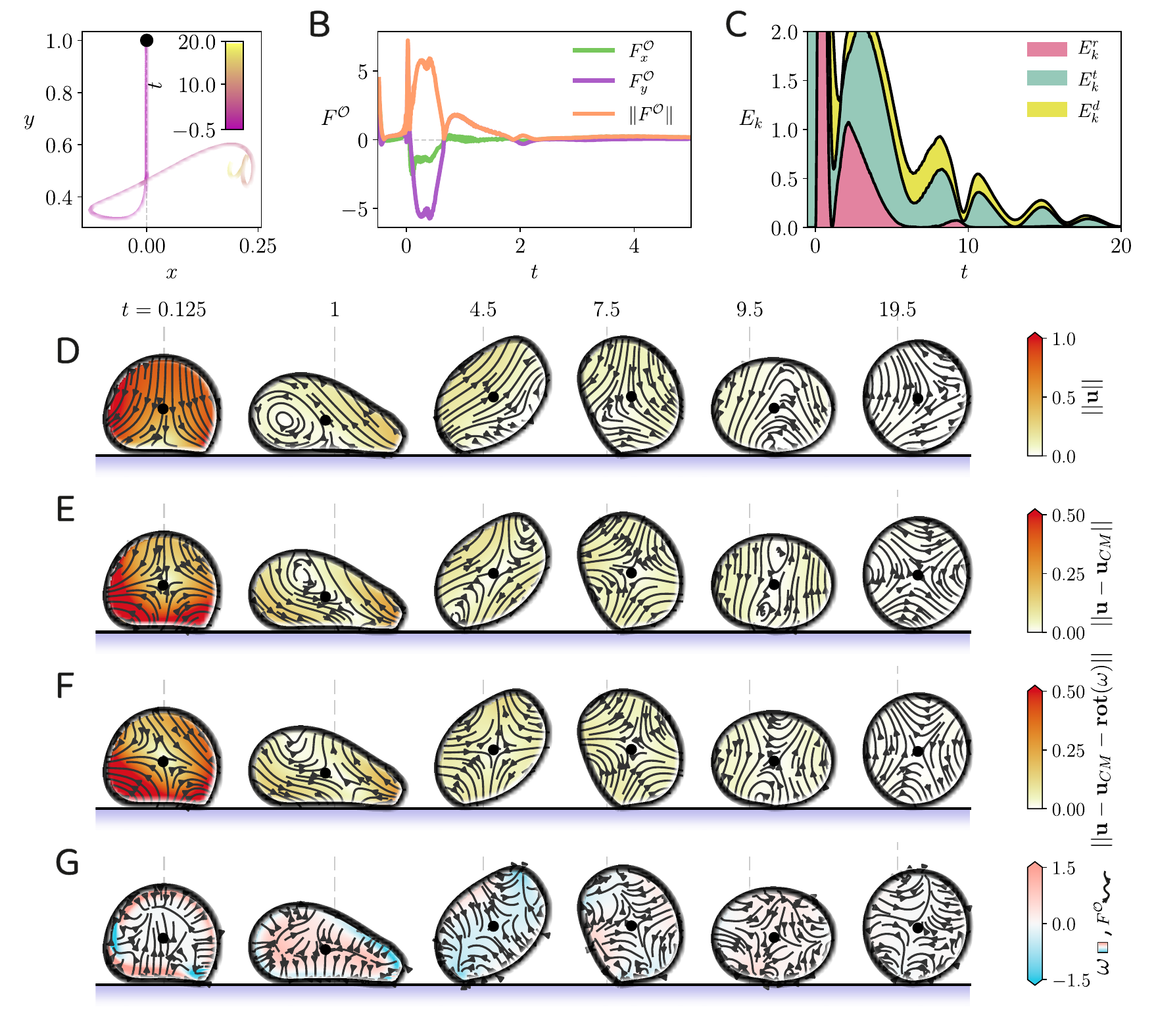}
\caption{\textbf{Same as above for a roller with moderate odd viscosity ($Re^\even = 50$, $Re^\odd = 4$, $We = 7$).} See \textbf{Supplementary Video VI} for a dynamic visualization of this simulation.}
\label{figAppendix:visualize_fields_slide_stop}
\end{figure}

Figure~\ref{figAppendix:visualize_fields_slide_stop} shows the velocity fields for an impact resulting in a droplet that slides over the surface and eventually stops. This case corresponds to $Re^\even=50$, $Re^\odd=4$, and $We=7$. Initially, the behavior is similar to that observed in previous figures: the droplet accelerates to the left during impact and then to the right during the spreading and retraction phases. The rightward acceleration is strong enough to visibly cause the droplet to slide to the right while still in contact with the surface. As shown by the kinetic energy distribution in Panel \textbf{C}, the leftward and rightward accelerations balance each other, leading to a rapid vanishing of the droplet's translational velocity. Although the kinetic energy continues to oscillate due to capillary-driven surface oscillations, it eventually dissipates through even viscosity, and the droplet comes to a near halt, reaching almost zero kinetic energy.

\subsection{Supplementary Videos}
\begin{enumerate}
\item \textbf{Video I:} Complementing Fig.~\ref{fig1:phenomenology}, we present the full dynamics of these two simulations (even and odd) in Supplementary Video I.
\item \textbf{Video II:} In Supplementary Video II, we present the full dynamics of the 7 simulations shown in Fig.~\ref{fig2:changing_Re_odd}. The value of $Re^{\odd}$ is changing for the fixed Weber number $We = 10$.
\item \textbf{Video III:} Supplementary Video III shows the full dynamics for the simulation presented in Fig. \ref{figAppendix:visualize_fields_bounce_left}. Panels \textbf{A} and \textbf{B} of this video contain, respectively, the average odd force and kinetic energy budget over time, as also shown in the original figure. Panel \textbf{C} shows the odd droplet as it moves and deforms through the domain during the impact process. In panel \textbf{D} the velocity field within the droplet is inspected. We show the total velocity field, as well as the residual velocities after subtraction of constant translation and solid body rotation. In panel \textbf{E} the vorticity within the droplet is presented, along with the components of the odd force $F^\odd = -\nabla \omega$. 
\item \textbf{Video IV:} Same as above, but corresponding to the simulation presented in Fig. \ref{figAppendix:visualize_fields_bounce_right}.
\item \textbf{Video V:} Same as above, but corresponding to the simulation presented in Fig. \ref{figAppendix:visualize_fields_roll}.
\item \textbf{Video VI:}  Same as above, but corresponding to the simulation presented in Fig. \ref{figAppendix:visualize_fields_slide_stop}.
\end{enumerate}

\pagebreak
\printbibliography


\end{document}

%% file: initialisation.tex
\usepackage{titlesec}
\titleformat{\section}
  {\bf\sffamily}
  {\thesection. }
  {5pt}
  {\MakeUppercase}
\renewcommand{\thesection}{\Roman{section}} 

\titleformat{\subsection}
  {\bf\sffamily}
  {\thesubsection. }
  {5pt}{}
  
\renewcommand{\thesubsection}{\Alph{subsection}}

\usepackage[affil-it]{authblk} 
\usepackage{etoolbox}
\usepackage{lmodern}

\usepackage{eurosym}

\usepackage[utf8]{inputenc}
\usepackage{geometry}
\usepackage[hidelinks]{hyperref}
\usepackage[citestyle=numeric,style=phys,biblabel=brackets,backend=bibtex,sorting=none,url=false,doi=false, natbib]{biblatex}
\bibliography{bibliography}
\DefineBibliographyStrings{english}{andothers={\itshape et\addabbrvspace al\adddot}}

\usepackage{csquotes}
\usepackage[]{authblk} 
\usepackage{etoolbox}
\usepackage{lmodern}

\usepackage{amsmath}
\usepackage{enumerate}
\usepackage{enumitem}
\usepackage{graphicx}
\usepackage{siunitx}
\usepackage{float}
\usepackage[]{parskip} 

\makeatletter
\patchcmd{\@maketitle}{\LARGE \@title}{\fontsize{16}{19.2}\selectfont\@title}{}{}
\makeatother

\usepackage[normalem]{ulem}

\usepackage{graphicx}
\usepackage{dcolumn}
\usepackage{bm}
\usepackage{tikz}
\usetikzlibrary{math}
\usetikzlibrary{shapes.geometric, arrows}
\usetikzlibrary{positioning}
\usetikzlibrary{shapes,arrows}
\usetikzlibrary{intersections}
\usepackage{csvsimple}
\usepackage{changepage}
\usepackage[tbtags]{mathtools}
\usepackage{siunitx}
\usepackage{csquotes}

\usepackage{float}
\usepackage{subfigure}
\usepackage[
	nonumberlist, 				
	acronym,      				
	nomain,						
	nopostdot,					
]{glossaries}
\usepackage{glossary-superragged}
\usepackage[hidelinks]{hyperref}
\usepackage{color, colortbl}

\usepackage{wrapfig}

\usepackage{pgfplots}
\usepackage{tikz}
\usetikzlibrary{arrows}
\usepackage{amsmath}
\pgfplotsset{compat=newest}
\usepgfplotslibrary{fillbetween}
\usetikzlibrary{calc}
\def\centerarc[#1](#2)(#3:#4:#5)
{ \draw[#1] ($(#2)+({#5*cos(#3)},{#5*sin(#3)})$) arc (#3:#4:#5); }

\usepackage{amssymb}

\usepackage{nicefrac}

\usepackage{abstract}

\usepackage{multirow}
\usepackage{tabularx}
\usepackage{booktabs}
\usepackage{array}
\usepackage{longtable}
\newcolumntype{L}[1]{>{\raggedright\let\newline\\\arraybackslash\hspace{0pt}}m{#1}}
\newcolumntype{C}[1]{>{\centering\let\newline\\\arraybackslash\hspace{0pt}}m{#1}}
\newcolumntype{R}[1]{>{\raggedleft\let\newline\\\arraybackslash\hspace{0pt}}m{#1}}

\newacronym{3d}{3D}{three dimensional}
\newacronym{am}{AM}{additive manufacturing}
\newacronym{fdm}{FDM}{fused deposition modeling}
\newacronym{ism}{ISM}{in-space manufacturing}
\newacronym{iss}{ISS}{International Space Station}
\newacronym{fcb}{FCB}{Functional Cargo Block}
\newacronym{dem}{DEM}{discrete element method}
\newacronym{md}{MD}{molecular dynamics}
\newacronym{dc}{DC}{direct-current}
\newacronym[plural=PFCs,firstplural=parabolic flight campaigns (PFCs)]{pfc}{PFC}{Parabolic Flight Campaign}
\newacronym{fft}{FFT}{Fast Fourrier Transform}
\newacronym{cad}{CAD}{Computer Assisted Design}
\newacronym{ptfe}{PTFE}{polytetrafluoroethylene}
\newacronym{ps}{PS}{polystyrene}
\newacronym{nasa}{NASA}{National Aeronautics and Space Administration}
\newacronym{esamm}{ESAMM}{Extended Structure Additive Manufacturing Machine}
\newacronym{amf}{AMF}{Additive Manufacturing Facility}
\newacronym{us}{US}{United States}
\newacronym{usa}{USA}{United States of America}
\newacronym{bmgs}{BMGs}{Bulk Metallic Glasses}
\newacronym{esa}{ESA}{European Space Agency}
\newacronym{si}{SI}{International System of Units, abbreviated from French \textit{Syst\`{e}me International (d'unit\'{e}s)}}
\newacronym{dlr}{DLR}{German Aerospace Center}
\newacronym{liggghts}{LIGGGHTS}{\acrshort{lammps} Improved for General Granular and Granular Heat Transfer Simulations}
\newacronym{lammps}{LAMMPS}{Large-scale Atomic/Molecular Massively Parallel Simulator}
\newacronym{sjkr}{SJKR}{Simplified Johnson-Kendall-Roberts}
\newacronym{ded}{DED}{Directed Energy Deposition}
\newacronym{slm}{SLM}{Selective Laser Melting}
\newacronym{sls}{SLS}{Selective Laser Sintering}
\newacronym{eva}{EVA}{Extra-Vehicular Activity}
\newacronym{sem}{SEM}{Scanning Electron Microscopy}
\newacronym{RPM}{RPM}{Ramdom Positioning Machine}
\newacronym{rpm}{rpm}{revolutions per minute}
\newacronym{rise}{RISE}{Research Internships in Science and Engineering}
\newacronym{daad}{DAAD}{German Academic Exchange Service, abbreviated from German \textit{Deutscher Akademischer Austauschdienst}}
\newacronym{fsm}{FSM}{finite-state machine}
\newacronym{ir}{IR}{infrared}
\newacronym{pcbs}{PCBs}{Printed Circuit Boards}
\newacronym{pcb}{PCB}{Printed Circuit Board}
\newacronym{mcr}{MCR}{Modular Compact Rheometer}
\newacronym{sff}{SFF}{Solid Freeform Fabrication}
\newacronym{uv}{UV}{ultraviolet}
\newacronym{abs}{ABS}{acrylonitrile butadiene styrene}
\newacronym{hpde}{HPDE}{high density polyethylene}
\newacronym{pei}{PEI}{polyetherimide}
\newacronym{bff}{BFF}{BioFabrication Facility}
\newacronym{lens}{LENS}{Laser Engineered Net Shaping}
\newacronym{cnc}{CNC}{Computer Numerical Control}
\newacronym{ebf3}{EBF$^3$}{Electron Beam Free-Form Fabrication}
\newacronym{leo}{LEO}{Low Earth Orbit}
\newacronym{pc}{PC}{polycarbonate}
\newacronym{crissp}{CRISSP}{Customisable Recyclable International Space Station Packaging}
\newacronym{Athena}{Athena}{Advanced Telescope for High-ENergy Astrophysics}
\newacronym{lbm}{LBM}{Laser Beam Melting}
\newacronym{bam}{BAM}{Federal Institute for Materials Research and Testing, abbreviated from German \textit{Bundesanstalt f\"{u}r Materialforschung und-pr\"{u}fung}}
\newacronym{pbf}{PBF}{powder bed fusion}
\newacronym{eb}{EB}{Electron Beam}
\newacronym{2d}{2D}{two dimensional}
\newacronym{4d}{4D}{four dimensional}
\newacronym{ft4}{FT4}{Freeman Technology 4 Powder Rheometer}
\newacronym{dsc}{DSC}{Differential Scanning Calorimetry}
\newacronym{pmma}{PMMA}{polymethylmethacrylate}
\newacronym{1g}{$1g$}{gravity on-ground}
\newacronym{mug}{$\mu g$}{microgravity}
\newacronym{bcm}{BCM}{Box Counting Method}
\newacronym{mct}{MCT}{Mode Coupling Theory}
\newacronym{gmct}{gMCT}{granular Mode Coupling Theory}
\newacronym{itt}{ITT}{Integration Through Transients}
\newacronym{mfc}{MFC}{Mass Flow Controller}
\newacronym{ct}{CT}{computed tomography}
\newacronym{xct}{XCT}{X-ray computed tomography}
\newacronym{cv}{CV}{curriculum vitae}
\newacronym{pi}{PI}{principal investigator}
\newacronym{osp}{OSP}{orthogonal superimposed perturbation}
\newacronym{npi}{NPI}{Network Partnering Initiative}
\newacronym{ecsat}{ECSAT}{European Centre for Space Applications and Telecommunications}
\newacronym{eac}{EAC}{European Astronaut Centre}
\newacronym{estec}{ESTEC}{European Space Research and Technology Centre}
\newacronym{fps}{fps}{frames per second}
\newacronym{pdf}{pdf}{probability density function}
\newacronym{al}{Al}{aluminium}
\newacronym{ss}{\textit{SS}}{\textit{Smooth Surface}}
\newacronym{rs}{\textit{RS}}{\textit{Rough Surface}}
\newacronym{rcp}{rcp}{random close packing}
\newacronym{iop}{IoP UvA}{Institute of Physics of the University of Amsterdam}
\newacronym{mp}{MP}{Institute of Material Physics for Space}
\newacronym{elgra}{ELGRA}{European Low Gravity Research Association}
\newacronym{zarm}{ZARM}{Center of Applied Space Technology and Microgravity}
\newacronym{piv}{PIV}{particle image velocimetry}

%% file: colors_define.tex
\definecolor{brickred}{rgb}{0.8, 0.25, 0.33}
\definecolor{darkorange}{rgb}{1.0, 0.55, 0.0}
\definecolor{persiangreen}{rgb}{0.0, 0.65, 0.58}
\definecolor{persianindigo}{rgb}{0.2, 0.07, 0.48}
\definecolor{cadet}{rgb}{0.33, 0.41, 0.47}
\definecolor{turquoisegreen}{rgb}{0.63, 0.84, 0.71}
\definecolor{sandybrown}{rgb}{0.96, 0.64, 0.38}
\definecolor{blueblue}{rgb}{0.0, 0.2, 0.6}
\definecolor{ballblue}{rgb}{0.13, 0.67, 0.8}
\definecolor{greengreen}{rgb}{0.0, 0.5, 0.0}

%% file: bibliography.bib
@article{avron1995viscosity,
  title={Viscosity of quantum Hall fluids},
  author={Avron, JE and Seiler, Ruedi and Zograf, Petr G},
  journal={Physical review letters},
  volume={75},
  number={4},
  pages={697},
  year={1995},
  publisher={APS}
}

@article{biance2006elasticity,
  title={On the elasticity of an inertial liquid shock},
  author={Biance, Anne-Laure and Chevy, Fr{\'e}d{\'e}ric and Clanet, Christophe and Lagubeau, Guillaume and Qu{\'e}r{\'e}, David},
  journal={Journal of Fluid Mechanics},
  volume={554},
  pages={47--66},
  year={2006},
  publisher={Cambridge University Press}
}

@article{souslov2020anisotropic,
  title={Anisotropic odd viscosity via a time-modulated drive},
  author={Souslov, Anton and Gromov, Andrey and Vitelli, Vincenzo},
  journal={Physical Review E},
  volume={101},
  number={5},
  pages={052606},
  year={2020},
  publisher={APS}
}

@article{lier2024odd,
  title={Odd viscous flow past a sphere at low but non-zero Reynolds numbers},
  author={Lier, Ruben},
  journal={Journal of Fluid Mechanics},
  volume={998},
  pages={A40},
  year={2024},
  publisher={Cambridge University Press}
}

@article{avron1998odd,
  title={Odd viscosity},
  author={Avron, JE},
  journal={Journal of statistical physics},
  volume={92},
  pages={543--557},
  year={1998},
  publisher={Springer}
}

@article{delacretaz2017transport,
  title={Transport signatures of the Hall viscosity},
  author={Delacr{\'e}taz, Luca V and Gromov, Andrey},
  journal={Physical review letters},
  volume={119},
  number={22},
  pages={226602},
  year={2017},
  publisher={APS}
}

@article{berdyugin2019measuring,
  title={Measuring Hall viscosity of graphene’s electron fluid},
  author={Berdyugin, Alexey I and Xu, SG and Pellegrino, FMD and Krishna Kumar, R and Principi, Alessandro and Torre, Iacopo and Ben Shalom, M and Taniguchi, Takashi and Watanabe, K and Grigorieva, IV and others},
  journal={Science},
  volume={364},
  number={6436},
  pages={162--165},
  year={2019},
  publisher={American Association for the Advancement of Science}
}

@article{scaffidi2017hydrodynamic,
  title={Hydrodynamic electron flow and Hall viscosity},
  author={Scaffidi, Thomas and Nandi, Nabhanila and Schmidt, Burkhard and Mackenzie, Andrew P and Moore, Joel E},
  journal={Physical review letters},
  volume={118},
  number={22},
  pages={226601},
  year={2017},
  publisher={APS}
}

@article{fritz2024hydrodynamic,
  title={Hydrodynamic electronic transport},
  author={Fritz, Lars and Scaffidi, Thomas},
  journal={Annual Review of Condensed Matter Physics},
  volume={15},
  number={1},
  pages={17--44},
  year={2024},
  publisher={Annual Reviews}
}

@article{markovich2021odd,
  title={Odd viscosity in active matter: Microscopic origin and 3D effects},
  author={Markovich, Tomer and Lubensky, Tom C},
  journal={Physical Review Letters},
  volume={127},
  number={4},
  pages={048001},
  year={2021},
  publisher={APS}
}

@article{furthauer2012active,
  title={Active chiral fluids},
  author={F{\"u}rthauer, S and Strempel, M and Grill, SW and J{\"u}licher, F},
  journal={The European Physical Journal E},
  volume={35},
  number={9},
  pages={89},
  year={2012},
  publisher={Springer-Verlag}
}

@article{soni2019odd,
  title={The odd free surface flows of a colloidal chiral fluid},
  author={Soni, Vishal and Bililign, Ephraim S and Magkiriadou, Sofia and Sacanna, Stefano and Bartolo, Denis and Shelley, Michael J and Irvine, William TM},
  journal={Nature physics},
  volume={15},
  number={11},
  pages={1188--1194},
  year={2019},
  publisher={Nature Publishing Group UK London}
}

@article{bililign2022motile,
  title={Motile dislocations knead odd crystals into whorls},
  author={Bililign, Ephraim S and Balboa Usabiaga, Florencio and Ganan, Yehuda A and Poncet, Alexis and Soni, Vishal and Magkiriadou, Sofia and Shelley, Michael J and Bartolo, Denis and Irvine, William TM},
  journal={Nature Physics},
  volume={18},
  number={2},
  pages={212--218},
  year={2022},
  publisher={Nature Publishing Group UK London}
}

@article{banerjee2017odd,
  title={Odd viscosity in chiral active fluids},
  author={Banerjee, Debarghya and Souslov, Anton and Abanov, Alexander G and Vitelli, Vincenzo},
  journal={Nature communications},
  volume={8},
  number={1},
  pages={1573},
  year={2017},
  publisher={Nature Publishing Group UK London}
}

@article{chen2024odd,
  title={Odd viscosity suppresses intermittency in direct turbulent cascades},
  author={Chen, Sihan and De Wit, Xander M and Fruchart, Michel and Toschi, Federico and Vitelli, Vincenzo},
  journal={Physical Review Letters},
  volume={133},
  number={14},
  pages={144002},
  year={2024},
  publisher={APS}
}

@article{de2024pattern,
  title={Pattern formation by turbulent cascades},
  author={de Wit, Xander M and Fruchart, Michel and Khain, Tali and Toschi, Federico and Vitelli, Vincenzo},
  journal={Nature},
  volume={627},
  number={8004},
  pages={515--521},
  year={2024},
  publisher={Nature Publishing Group UK London}
}

@article{souslov2019topological,
  title={Topological waves in fluids with odd viscosity},
  author={Souslov, Anton and Dasbiswas, Kinjal and Fruchart, Michel and Vaikuntanathan, Suriyanarayanan and Vitelli, Vincenzo},
  journal={Physical review letters},
  volume={122},
  number={12},
  pages={128001},
  year={2019},
  publisher={APS}
}

@article{lier2023lift,
  title={Lift force in odd compressible fluids},
  author={Lier, Ruben and Duclut, Charlie and Bo, Stefano and Armas, Jay and J{\"u}licher, Frank and Sur{\'o}wka, Piotr},
  journal={Physical Review E},
  volume={108},
  number={2},
  pages={L023101},
  year={2023},
  publisher={APS}
}

@article{khain2022stokes,
  title={Stokes flows in three-dimensional fluids with odd and parity-violating viscosities},
  author={Khain, Tali and Scheibner, Colin and Fruchart, Michel and Vitelli, Vincenzo},
  journal={Journal of Fluid Mechanics},
  volume={934},
  pages={A23},
  year={2022},
  publisher={Cambridge University Press}
}

@article{aggarwal2023thermocapillary,
  title={Thermocapillary migrating odd viscous droplets},
  author={Aggarwal, Aaveg and Kirkinis, Eleftherios and Olvera de la Cruz, Monica},
  journal={Physical review letters},
  volume={131},
  number={19},
  pages={198201},
  year={2023},
  publisher={APS}
}

@article{hosaka2021hydrodynamic,
  title={Hydrodynamic lift of a two-dimensional liquid domain with odd viscosity},
  author={Hosaka, Yuto and Komura, Shigeyuki and Andelman, David},
  journal={Physical Review E},
  volume={104},
  number={6},
  pages={064613},
  year={2021},
  publisher={APS}
}

@article{naraigh2023analysis,
  title={Analysis of the spreading radius in droplet impact: The two-dimensional case},
  author={N{\'a}raigh, Lennon {\'O} and Mairal, Juan},
  journal={Physics of Fluids},
  volume={35},
  number={10},
  year={2023},
  publisher={AIP Publishing}
}

@article{poggioli2023odd,
  title={Odd mobility of a passive tracer in a chiral active fluid},
  author={Poggioli, Anthony R and Limmer, David T},
  journal={Physical Review Letters},
  volume={130},
  number={15},
  pages={158201},
  year={2023},
  publisher={APS}
}

@article{samanta2022role,
  title={Role of odd viscosity in falling viscous fluid},
  author={Samanta, Arghya},
  journal={Journal of Fluid Mechanics},
  volume={938},
  pages={A9},
  year={2022},
  publisher={Cambridge University Press}
}

@article{abanov2019free,
  title={Free-surface variational principle for an incompressible fluid with odd viscosity},
  author={Abanov, Alexander G and Monteiro, Gustavo M},
  journal={Physical review letters},
  volume={122},
  number={15},
  pages={154501},
  year={2019},
  publisher={APS}
}

@article{abanov2018odd,
  title={Odd surface waves in two-dimensional incompressible fluids},
  author={Abanov, Alexander G and Can, Tankut and Ganeshan, Sriram},
  journal={SciPost Physics},
  volume={5},
  number={1},
  pages={010},
  year={2018}
}

@article{kirkinis2019odd,
  title={Odd-viscosity-induced stabilization of viscous thin liquid films},
  author={Kirkinis, E and Andreev, AV},
  journal={Journal of Fluid Mechanics},
  volume={878},
  pages={169--189},
  year={2019},
  publisher={Cambridge University Press}
}

@article{josserand2016drop,
  title={Drop impact on a solid surface},
  author={Josserand, Christophe and Thoroddsen, Sigurdur T},
  journal={Annual review of fluid mechanics},
  volume={48},
  number={1},
  pages={365--391},
  year={2016},
  publisher={Annual Reviews}
}

@article{yarin2006drop,
  title={Drop impact dynamics: splashing, spreading, receding, bouncing…},
  author={Yarin, Alexander L},
  journal={Annu. Rev. Fluid Mech.},
  volume={38},
  number={1},
  pages={159--192},
  year={2006},
  publisher={Annual Reviews}
}

@article{sanjay2024unifying,
  title={Unifying theory of scaling in drop impact: Forces \& maximum spreading diameter},
  author={Sanjay, Vatsal and Lohse, Detlef},
  journal={arXiv preprint arXiv:2408.12714},
  year={2024}
}

@article{quere2005non,
  title={Non-sticking drops},
  author={Qu{\'e}r{\'e}, David},
  journal={Reports on Progress in Physics},
  volume={68},
  number={11},
  pages={2495},
  year={2005},
  publisher={IOP Publishing}
}

@article{pasandideh1996capillary,
  title={Capillary effects during droplet impact on a solid surface},
  author={Pasandideh-Fard, M and Qiao, YM and Chandra, Sanjeev and Mostaghimi, Javad},
  journal={Physics of fluids},
  volume={8},
  number={3},
  pages={650--659},
  year={1996},
  publisher={American Institute of Physics}
}

@article{wildeman2016spreading,
  title={On the spreading of impacting drops},
  author={Wildeman, Sander and Visser, Claas Willem and Sun, Chao and Lohse, Detlef},
  journal={Journal of fluid mechanics},
  volume={805},
  pages={636--655},
  year={2016},
  publisher={Cambridge University Press}
}

@article{clanet2004maximal,
  title={Maximal deformation of an impacting drop},
  author={Clanet, Christophe and B{\'e}guin, C{\'e}dric and Richard, Denis and Qu{\'e}r{\'e}, David},
  journal={Journal of Fluid Mechanics},
  volume={517},
  pages={199--208},
  year={2004},
  publisher={Cambridge University Press}
}

@article{zhang2022impact,
  title={Impact forces of water drops falling on superhydrophobic surfaces},
  author={Zhang, Bin and Sanjay, Vatsal and Shi, Songlin and Zhao, Yinggang and Lv, Cunjing and Feng, Xi-Qiao and Lohse, Detlef},
  journal={Physical review letters},
  volume={129},
  number={10},
  pages={104501},
  year={2022},
  publisher={APS}
}

@article{Joanes1998-Skewness,
 ISSN = {00390526, 14679884},
 URL = {http://www.jstor.org/stable/2988433},
 author = {D. N. Joanes and C. A. Gill},
 journal = {Journal of the Royal Statistical Society. Series D (The Statistician)},
 number = {1},
 pages = {183--189},
 publisher = {[Royal Statistical Society, Wiley]},
 title = {Comparing Measures of Sample Skewness and Kurtosis},
 urldate = {2025-01-17},
 volume = {47},
 year = {1998}
}

@article{Fruchart2023,
  title = {Odd Viscosity and Odd Elasticity},
  volume = {14},
  ISSN = {1947-5462},
  url = {http://dx.doi.org/10.1146/annurev-conmatphys-040821-125506},
  DOI = {10.1146/annurev-conmatphys-040821-125506},
  number = {1},
  journal = {Annual Review of Condensed Matter Physics},
  publisher = {Annual Reviews},
  author = {Fruchart,  Michel and Scheibner,  Colin and Vitelli,  Vincenzo},
  year = {2023},
  month = mar,
  pages = {471–510}
}

@article{tauber2020anomalous,
  title={Anomalous bulk-edge correspondence in continuous media},
  author={Tauber, Cl{\'e}ment and Delplace, Pierre and Venaille, Antoine},
  journal={Physical Review Research},
  volume={2},
  number={1},
  pages={013147},
  year={2020},
  publisher={APS}
}

@article{urzhumov2011fluid,
  title={Fluid flow control with transformation media},
  author={Urzhumov, Yaroslav A and Smith, David R},
  journal={Physical review letters},
  volume={107},
  number={7},
  pages={074501},
  year={2011},
  publisher={APS}
}

@article{park2019hydrodynamic,
  title={Hydrodynamic metamaterial cloak for drag-free flow},
  author={Park, Juhyuk and Youn, Jae Ryoun and Song, Young Seok},
  journal={Physical review letters},
  volume={123},
  number={7},
  pages={074502},
  year={2019},
  publisher={APS}
}

@article{djellouli2024shell,
  title={Shell buckling for programmable metafluids},
  author={Djellouli, Adel and Van Raemdonck, Bert and Wang, Yang and Yang, Yi and Caillaud, Anthony and Weitz, David and Rubinstein, Shmuel and Gorissen, Benjamin and Bertoldi, Katia},
  journal={Nature},
  volume={628},
  number={8008},
  pages={545--550},
  year={2024},
  publisher={Nature Publishing Group UK London}
}

@article{ewoldt2022designing,
  title={Designing complex fluids},
  author={Ewoldt, Randy H and Saengow, Chaimongkol},
  journal={Annual Review of Fluid Mechanics},
  volume={54},
  number={1},
  pages={413--441},
  year={2022},
  publisher={Annual Reviews}
}

@misc{Popinet2013-Basilisk,   
    title = {Basilisk C},   
    url = {http://basilisk.fr},   
    author = {S. Popinet and Collaborators},   
    year = {2013-2021},   
    note = {Accessed on 01 01, 2024} 
}

@article{popinet2009accurate,
  title={An accurate adaptive solver for surface-tension-driven interfacial flows},
  author={Popinet, St{\'e}phane},
  journal={Journal of Computational Physics},
  volume={228},
  number={16},
  pages={5838--5866},
  year={2009},
  publisher={Elsevier}
}

@article{Popinet2015,
  doi = {10.1016/j.jcp.2015.09.009},
  url = {https://doi.org/10.1016/j.jcp.2015.09.009},
  year = {2015},
  month = dec,
  publisher = {Elsevier {BV}},
  volume = {302},
  pages = {336--358},
  author = {St{\'{e}}phane Popinet},
  title = {A quadtree-adaptive multigrid solver for the Serre{\textendash}Green{\textendash}Naghdi equations},
  journal = {Journal of Computational Physics}
}

@article{Hirt1981,
  doi = {10.1016/0021-9991(81)90145-5},
  url = {https://doi.org/10.1016/0021-9991(81)90145-5},
  year = {1981},
  month = jan,
  publisher = {Elsevier {BV}},
  volume = {39},
  number = {1},
  pages = {201--225},
  author = {C.W Hirt and B.D Nichols},
  title = {Volume of fluid ({VOF}) method for the dynamics of free boundaries},
  journal = {Journal of Computational Physics}
}

@BOOK{Tryggvason2011-book,
  title     = "Direct numerical simulations of gas-liquid multiphase flows",
  author    = "Tryggvason, Gretar and Scardovelli, Ruben and Zaleski, Stephane",
  publisher = "Cambridge University Press",
  month     =  mar,
  year      =  2011,
  address   = "Cambridge, England"
}

@article{jalaal2012fragmentation,
  title={Fragmentation of falling liquid droplets in bag breakup mode},
  author={Jalaal, M and Mehravaran, K},
  journal={International Journal of Multiphase Flow},
  volume={47},
  pages={115--132},
  year={2012},
  publisher={Elsevier}
}

@article{francca2024elasto,
  title={Elasto-viscoplastic spreading: From plastocapillarity to elastocapillarity},
  author={Fran{\c{c}}a, Hugo L and Jalaal, Maziyar and Oishi, Cassio M},
  journal={Physical Review Research},
  volume={6},
  number={1},
  pages={013226},
  year={2024},
  publisher={APS}
}

@article{jalaal2021spreading,
  title={Spreading of viscoplastic droplets},
  author={Jalaal, Maziyar and Stoeber, Boris and Balmforth, Neil J},
  journal={Journal of Fluid Mechanics},
  volume={914},
  pages={A21},
  year={2021},
  publisher={Cambridge University Press}
}

@article{cimpeanu2018early,
  title={Early-time jet formation in liquid--liquid impact problems: theory and simulations},
  author={Cimpeanu, Radu and Moore, Madeleine Rose},
  journal={Journal of Fluid Mechanics},
  volume={856},
  pages={764--796},
  year={2018},
  publisher={Cambridge University Press}
}

@article{sanjay2025role,
  title={The role of viscosity on drop impact forces on non-wetting surfaces},
  author={Sanjay, Vatsal and Zhang, Bin and Lv, Cunjing and Lohse, Detlef},
  journal={Journal of Fluid Mechanics},
  volume={1004},
  pages={A6},
  year={2025},
  publisher={Cambridge University Press}
}

@article{jalaal2018gel,
  title={Gel-controlled droplet spreading},
  author={Jalaal, M and Seyfert, C and Stoeber, B and Balmforth, NJ},
  journal={Journal of Fluid Mechanics},
  volume={837},
  pages={115--128},
  year={2018},
  publisher={Cambridge University Press}
}

@article{jalaal2014controlled,
  title={Controlled spreading of thermo-responsive droplets},
  author={Jalaal, Maziyar and Stoeber, Boris},
  journal={Soft Matter},
  volume={10},
  number={6},
  pages={808--812},
  year={2014},
  publisher={Royal Society of Chemistry}
}

@article{Veenstra2025Adaptive,
  title={Adaptive locomotion of active solids},
  author={Veenstra, Jonas and Scheibner,  Colin and Brandenbourger, Martin and Jack, Binysh and Souslov,Anton and Vitelli, Vincenzo and Coulais, Corentin},
  journal={Nature},
  volume={},
  number={},
  pages={},
  year={2025}
}

@article{lier2022passive,
  title={Passive odd viscoelasticity},
  author={Lier, Ruben and Armas, Jay and Bo, Stefano and Duclut, Charlie and J{\"u}licher, Frank and Sur{\'o}wka, Piotr},
  journal={Physical Review E},
  volume={105},
  number={5},
  pages={054607},
  year={2022},
  publisher={APS}
}

@article{banerjee2021active,
  title={Active viscoelasticity of odd materials},
  author={Banerjee, Debarghya and Vitelli, Vincenzo and J{\"u}licher, Frank and Sur{\'o}wka, Piotr},
  journal={Physical Review Letters},
  volume={126},
  number={13},
  pages={138001},
  year={2021},
  publisher={APS}
}

@article{lohse2022fundamental,
  title={Fundamental fluid dynamics challenges in inkjet printing},
  author={Lohse, Detlef},
  journal={Annual review of fluid mechanics},
  volume={54},
  number={1},
  pages={349--382},
  year={2022},
  publisher={Annual Reviews}
}

@article{ahmed2018maximum,
  title={Maximum spreading of a ferrofluid droplet under the effect of magnetic field},
  author={Ahmed, Abrar and Fleck, Brian A and Waghmare, Prashant R},
  journal={Physics of Fluids},
  volume={30},
  number={7},
  year={2018},
  publisher={AIP Publishing}
}

@article{jezervsek2023control,
  title={Control of droplet impact through magnetic actuation of surface microstructures},
  author={Jezer{\v{s}}ek, Matija and Kriegl, Raphael and Kravanja, Gaia and Hribar, Luka and Dreven{\v{s}}ek-Olenik, Irena and Unold, Heiko and Shamonin, Mikhail},
  journal={Advanced Materials Interfaces},
  volume={10},
  number={11},
  pages={2202471},
  year={2023},
  publisher={Wiley Online Library}
}

@article{yang2022droplet,
  title={Droplet manipulation on superhydrophobic surfaces based on external stimulation: A review},
  author={Yang, Chen and Zeng, Qinghong and Huang, Jinxia and Guo, Zhiguang},
  journal={Advances in Colloid and interface Science},
  volume={306},
  pages={102724},
  year={2022},
  publisher={Elsevier}
}

@article{zhao2023regulating,
  title={Regulating droplet impact symmetry by surface engineering},
  author={Zhao, Zhipeng and Li, Huizeng and Liu, Quan and Li, An and Xue, Luanluan and Yuan, Renxuan and Yu, Xinye and Li, Rujun and Deng, Xiao and Song, Yanlin},
  journal={Droplet},
  volume={2},
  number={2},
  pages={e52},
  year={2023},
  publisher={Wiley Online Library}
}

@article{cosme2023nonlinear,
  title={Nonlinear density waves on graphene electron fluids},
  author={Cosme, Pedro and Ter{\c{c}}as, Hugo},
  journal={Physical Review B},
  volume={107},
  number={19},
  pages={195432},
  year={2023},
  publisher={APS}
}

@article{ramesh2025frozen,
  title={Frozen by Heating: Temperature Controlled Dynamic States in Droplet Microswimmers},
  author={Ramesh, Prashanth and Chen, Yibo and R{\"a}der, Petra and Morsbach, Svenja and Jalaal, Maziyar and Maass, Corinna C},
  journal={Advanced Materials},
  pages={2416813},
  publisher={Wiley Online Library}
}
